\documentclass[aps,prd,preprint,superscriptaddress,tightenlines,%
nofootinbib]{revtex4}
\newcommand{\PRE}[1]{{#1}}   

\usepackage{epsfig}

\newcommand{\postscript}[2]{\setlength{\epsfxsize}{#2\hsize}
   \centerline{\epsfbox{#1}}}

\newcommand{\mplanck}{M_{\text{Pl}}}
\newcommand{\mstar}{M_{\ast}}
\newcommand{\md}{M_D}
\newcommand{\mbh}{M_{\text{BH}}}
\newcommand{\mbhmin}{M_{\text{BH}}^{\text{min}}}

\newcommand{\xmin}{x_{\text{min}}}

\newcommand{\ev}{\text{eV}}

\newcommand{\gev}{\text{GeV}}
\newcommand{\tev}{\text{TeV}}
\newcommand{\pb}{\text{pb}}
\newcommand{\cm}{\text{cm}}
\newcommand{\km}{\text{km}}
\newcommand{\g}{\text{g}}
\newcommand{\s}{\text{s}}

\newcommand{\sr}{\text{sr}}

\newcommand{\kmwe}{\text{kmwe}}
\newcommand{\xmax}{X_{\text{max}}}
\newcommand{\etal}{{\em et al.}}
\newcommand{\eg}{{\em e.g.}}

\newcommand{\eqref}[1]{Eq.~(\ref{#1})}

\begin{document}

\preprint{
\hfil
\begin{minipage}[t]{3in}
\begin{flushright}
\vspace*{.4in}
NUB--3224--Th--01\\
MIT--CTP--3216\\
UCI--TR--2001--42\\
UK/01--12\\
hep-ph/0112247
\end{flushright}
\end{minipage}
}

\title{
\PRE{\vspace*{1.5in}}
Black Holes from Cosmic Rays: Probes of Extra Dimensions and
New Limits on TeV-Scale Gravity

\PRE{\vspace*{0.3in}}
}

\author{Luis A.~Anchordoqui}
\affiliation{Department of Physics,\\
Northeastern University, Boston, MA 02115
\PRE{\vspace*{.1in}}
}

\author{Jonathan L.~Feng}
\affiliation{Center for Theoretical Physics,\\
Massachusetts Institute of Technology, Cambridge, MA 02139
\PRE{\vspace*{.1in}}
}
\affiliation{Department of Physics and Astronomy,\\
University of California, Irvine, CA 92697
\PRE{\vspace*{.1in}}
}

\author{Haim Goldberg}
\affiliation{Department of Physics,\\
Northeastern University, Boston, MA 02115
\PRE{\vspace*{.1in}}
}

\author{Alfred D.~Shapere}%
\affiliation{Department of Physics,\\
University of Kentucky, Lexington, KY 40502
\PRE{\vspace*{.5in}}
}


\begin{abstract}
\PRE{\vspace*{.1in}}
If extra spacetime dimensions and low-scale gravity exist, black holes
will be produced in observable collisions of elementary particles.
For the next several years, ultra-high energy cosmic rays provide the
most promising window on this phenomenon.  In particular, cosmic
neutrinos can produce black holes deep in the Earth's atmosphere,
leading to quasi-horizontal giant air showers.  We determine the
sensitivity of cosmic ray detectors to black hole production and
compare the results to other probes of extra dimensions.  With $n \ge
4$ extra dimensions, current bounds on deeply penetrating showers from
AGASA already provide the most stringent bound on low-scale gravity,
requiring a fundamental Planck scale $M_D > 1.3 - 1.8~\tev$.  The
Auger Observatory will probe $M_D$ as large as 4 TeV and may observe
on the order of a hundred black holes in 5 years.  We also consider
the implications of angular momentum and possible exponentially
suppressed parton cross sections; including these effects, large black
hole rates are still possible.  Finally, we demonstrate that even if
only a few black hole events are observed, a standard model
interpretation may be excluded by comparison with Earth-skimming
neutrino rates.
\end{abstract}

\pacs{04.70.-s, 96.40.Tv, 13.15.+g, 04.50.+h}

\maketitle

\section{Introduction}

Tiny black holes (BHs) can be produced in particle collisions with
center-of-mass energies above the fundamental scale of
gravity~\cite{Amati:1987wq,'tHooft:rb}, where they should be
well-described semi-classically and
thermodynamically~\cite{Hawking:1975sw}. In conventional 4-dimensional
theories, {\em viz.}, where the Planck scale $\sim 10^{19}~\gev$ is
fundamental and the weak scale $\sim 1$ TeV is derived from it via
some dynamical mechanism, the study of such BHs is far beyond the
realm of experimental particle physics. Over the last few years,
however, physicists have begun exploring an alternative approach to
the longstanding gauge hierarchy problem, wherein the weak scale
becomes the fundamental scale of nature and the Planck scale is
derived from this, with the hierarchy in scales a consequence of large
or warped extra dimensions~\cite{Antoniadis:1990ew,Randall:1999ee}. If
this is the case, the fundamental scale of gravity can be ${\cal
O}$(TeV), and BH production and evaporation may be observed in
collisions of elementary particles~\cite{Banks:1999gd,Emparan:2000rs,%
Giddings:2000ay,Giddings:2001bu,Dimopoulos:2001hw}.

If gravity indeed becomes strong at the TeV scale, ultra-high energy
cosmic rays provide a powerful opportunity to probe BH production at
super-Planckian energies~\cite{Feng:2001ib}.  Cosmic rays with
energies $\sim 10^{19}~\ev$ have been observed~\cite{Nagano:ve}.  They
interact in the Earth's atmosphere and crust with center-of-mass
energies $\sim 100~\tev$, far beyond the reach of present and planned
man-made colliders.  These cosmic rays may therefore produce BHs,
allowing cosmic ray detectors to test the existence of TeV-scale
gravity and extra dimensions by searching for evidence of BH
production~\cite{Feng:2001ib,Anchordoqui:2001ei,%
Emparan:2001kf,Ringwald:2001vk,Uehara:2001yk}. A particularly
promising signal is provided by ultra-high energy cosmic neutrinos,
which may produce BHs with cross sections two or more orders of
magnitude above their standard model (SM) interactions.  These BHs
will decay promptly in a thermal distribution of SM particles.  Of the
order of a hundred BH events may be detected at the Auger
Observatory~\cite{Feng:2001ib} as quasi-horizontal, deeply penetrating
showers with distinctive properties~\cite{Anchordoqui:2001ei}.  The
possibility of BH production by cosmic rays supplements possible
sub-Planckian signatures of low-scale
gravity~\cite{Nussinov:1999jt,Jain:2000pu,Tyler:2001gt,%
Alvarez-Muniz:2001mk,Sigl:2001th}.

In this article we extend previous work to derive bounds from the
non-observation of BH-initiated showers in current data at the Akeno
Giant Air Shower Array (AGASA).  We also extend previous analyses of
BH discovery prospects at Auger, and discuss in detail the possibility
of distinguishing BH events from SM events.  A preliminary version of
some of these results was presented in Ref.~\cite{GAP}.

We begin in Sec.~\ref{sec:limits} with an overview of TeV-scale
gravity.  We collect and review existing bounds on the fundamental
Planck scale in a uniform convention.  In Sec.~\ref{sec:BH} we discuss
semiclassical BH production, including the effects of angular momentum
and the production of Kerr BHs, as well as the proposed exponential
suppression advocated by
Voloshin~\cite{Voloshin:2001vs,Voloshin:2001fe}.  This is followed in
Secs.~\ref{sec:flux} and \ref{sec:acceptance} by detailed discussions
of cosmogenic neutrino fluxes and ground array experiments,
respectively.

Our results for event rates and new limits on the scale of
higher-dimensional gravity are presented in Secs.~\ref{sec:AGASA} and
\ref{sec:Auger}.  We begin with current data from AGASA.  The AGASA
Collaboration has already reported no significant signal for neutrino
air showers during an observation time (live) of 1710.5
days~\cite{agasa}. Given the standard assumption of a geometric black
hole cross section, we find that this data implies the most stringent
bound on the fundamental Planck scale to date for $n \ge 4$ extra
dimensions, exceeding limits derived~\cite{Ringwald:2001vk} from Fly's
Eye data~\cite{Baltrusaitis:mt} and also more stringent than the
constraints from graviton emission and exchange obtained by the
LEP~\cite{Pagliarone:2001ff} and D\O~\cite{Abbott:2000zb}
Collaborations.  In Sec.~\ref{sec:Auger} we then consider the
prospects for BH production at the Auger Observatory.  Tens of black
holes may be observed per year; conversely, non-observation of BHs
will imply bounds as large as 4 TeV on the fundamental Planck scale.

In Sec.~\ref{sec:skimming} we note that comparison to Earth-skimming
neutrino event
rates~\cite{Bertou:2001vm,Feng:2001ue,Kusenko:2001gj,Domokos:1997ve}
allows one to distinguish BH events from SM events. This point was
noted already in Ref.~\cite{Feng:2001ib}, but was not considered in
Ref.~\cite{Ringwald:2001vk}, leading to weaker conclusions.  Here, we
consider this point quantitatively and find that, even with a handful
of BH events, a SM explanation may be excluded based on event rates
alone.  If seen, black holes created by cosmic rays will provide the
first evidence for extra dimensions and TeV-scale gravity, initiating
an era of detailed study of black hole properties at both cosmic ray
detectors and future colliders, such as the
LHC~\cite{Giddings:2001bu,Dimopoulos:2001hw,Cheung:2001ue,%
Park:2001xc,Rizzo:2001dk,Dimopoulos:2001qe,Hossenfelder:2001dn}.  Our
conclusions are collected in Sec.~\ref{sec:conclusions}.

\section{Existing limits on low-scale gravity}
\label{sec:limits}

Depending on the dimensionality and the particular form of spacetime,
the gauge hierarchy problem may be re-expressed as a hierarchy in
length scales. In the canonical example~\cite{Antoniadis:1990ew},
spacetime is a direct product of a non-compact 4-dimensional spacetime
manifold and a flat spatial $n$-torus of common linear size $2\pi r_c$
and volume $V_n = (2\pi r_c)^n$. Only gravity propagates in the full
$(4{+}n)$-dimensional spacetime; all others fields are confined to a
3-brane extended in the non-compact dimensions. Here, the low energy
4-dimensional Planck scale $\mplanck$ is related to the fundamental
scale of gravity in $(4{+}n)$ dimensions, $M_*$, according to
\begin{equation}
\mplanck^2 = M_*^{2+n} V_n = V_n/G_{(4+n)}\ \ ,
\label{mpl}
\end{equation}
with $G_{(4+n)}$ defined by the $(4+n)$ dimensional Einstein field
equation $R_{AB}-\frac{1}{2}g_{AB}=-8\pi G_{(4+n)}\ T_{AB}$. In what
follows it will be convenient to work with the mass
scale~\cite{Giudice:1998ck}
\begin{equation}
\md = [(2\pi)^n/8\pi]^{1/(n+2)}\, \mstar \ , \quad D=4+n\ .
\label{mdmstar}
\end{equation}
If $r_c$ is significantly larger than the Planck length, a hierarchy
is introduced between $\mplanck$ and $\md$, and gravity becomes strong
in the entire $(4+n)$-dimensional spacetime at the scale $\md$ far
below the conventional Planck scale $\mplanck \sim 10^{19}~\gev$. Our
conclusions will be essentially unchanged for more general
``asymmetric'' compactifications, with, \eg, $p$ ``small'' dimensions
with sizes $\alt\tev^{-1}$ and $n_{\rm eff} = n-p$ large extra
dimensions~\cite{Lykken:1999ms}. (Note, however, that in this
case, the production of brane configurations wrapped around small
extra dimensions may be competitive with black hole
production~\cite{Ahn:2002mj}.)  Many of our results for black hole
production and detection also apply for warped
compactifications~\cite{Randall:1999ee} in which the curvature length
is much larger than a $\tev^{-1}$. Hereafter, we will focus our
discussion on bounds in flat compactification scenarios. In the
figures, for $n=1$ results, warped scenarios are implicit.

\subsection{Bounds from Newtonian gravity}
\label{sec:Newtonian}

The provocative new features of these scenarios have motivated many
phenomenological studies to assess their experimental viability.
Naturally, the most obvious consequence of the existence of large
compact dimensions is the deviation from Newtonian gravity at
distances of order $r_c$.  For $n=1$ and $\md \sim 1~\tev$, $r_c \sim
10^{13}~\cm$, implying deviations from Newtonian gravity over solar
system distances, so this case is empirically excluded. For $n=2$,
sub-millimeter tests of the gravitational inverse-square law constrain
$\md > 1.6~\tev$~\cite{Hoyle:2000cv}. For $n \geq 3$, $r_c$ becomes
microscopic and therefore eludes the search for deviations in
gravitational measurements.

\subsection{Astrophysical bounds}
\label{sec:astrophysical}

In the presence of large compact dimensions, however, the effects of
gravity are enhanced at high energies, due to the accessibility of
numerous excited states of the graviton (referred to as Kaluza-Klein
(KK) gravitons~\cite{Kaluza:tu}), corresponding to excitations of the
graviton field in the compactified dimensions. For low numbers of
extra dimensions, by far the most restrictive limits on the radii of
large compact dimensions come from the effects of KK graviton emission
on cooling of supernovae, and from neutron star heating by KK
decays~\cite{Cullen:1999hc}.  For $n=2$ the latter requires $\md > 600
- 1800~\tev$, far above the weak scale; for $n=3$, the bound is $\md >
10 - 100~\tev$. These limits apply only for the situation where all
extra dimensions have the same compactification radius. In the general
case, the bounds could be less restrictive.

\subsection{Collider bounds}
\label{sec:tevatron}

For $n \geq 4$ extra dimensions, only high energy collisions are
useful as probes.  The effects of direct graviton emission, including
production of single photons or $Z$'s, were sought at
LEP~\cite{Acciarri:1999jy}.  The resulting bounds are fairly
model-independent, as the relatively low energies at LEP imply a
negligible dependence on the soft-brane damping factor discussed
below.  For $n=4\ (6)$, these null results imply $M_D > 870\
(610)~\gev$~\cite{Pagliarone:2001ff}.

The effects of low-scale gravity can also be seen through virtual
graviton effects.  These are most stringently bounded by the D\O\
Collaboration, which recently reported~\cite{Abbott:2000zb} the first
results for virtual graviton effects at a hadron collider.  The data
collected at $\sqrt{s} = 1.8~\tev$ for dielectron and diphoton
production at the Tevatron agree well with the SM predictions and
provide the most restrictive limits on an {\em effective}
extra-dimensional Planck scale for $n\geq 4$. This scale (called
$\Lambda_T$ in~\cite{Giudice:1998ck}, and related to $M_S$
in~\cite{Han:1998sg,Hewett:1998sn}) simply parameterizes the KK
graviton exchange amplitudes for these processes: except for the
different conventions used, they simply convey the experimental limit
in terms of an energy-independent four-point function.  In the context
of low-scale gravity, the effective scale depends on both $G_{(4+n)}$
and on an ultraviolet cutoff on the contributing KK
modes~\cite{Giudice:1998ck,Han:1998sg,Hewett:1998sn}. This cutoff
represents the energy where emission of graviton modes from the brane
into the extra dimensions are damped by the effects of a non-rigid
brane, and it is expected to be of order $G_{(4+n)}^{-1/(n+2)}$.

In this work we will use a Gaussian
cutoff~\cite{Bando:1999di,Bando:2000ch}, which emerges if one includes
in the interaction the brane Goldstone modes.  With this cutoff, a
form factor $e^{-m^2/2\Lambda^2}$ is introduced at each
graviton-matter vertex, where $m$ is the mass of the graviton and
$\Lambda$ parameterizes the cutoff. In real graviton emission
processes, the effect of the cutoff is somewhat alleviated because of
finite cuts on the missing energy.  However, at LHC energies, the
corrections become significant~\cite{Murayama:2001av} in the expected
region $\Lambda < M_D$, and are of order 100\% when $\Lambda/M_D
\simeq 0.5$.  For virtual processes, the $s$-channel diphoton or
dielectron amplitude has the form~\cite{Giudice:1998ck,Han:1998sg}
\begin{eqnarray}
{\cal A}&= &{\cal S}(\hat s)\,{\cal T}\nonumber\\
{\cal S}(\hat s)&=&\frac{8\pi}{\mplanck^2}\sum_{\vec \ell}\,
\frac{1}{m^2-\hat s}\nonumber\\
{\cal T}&=& T_{\mu\nu}T^{\mu\nu}-\frac{1}{n+2}
T^{\mu}_{\mu} T^{\nu}_{\nu} \ ,
\label{as}
\end{eqnarray}
where the sum on $\vec{\ell}$ denotes a sum over the KK graviton
modes, labeled by an $n$--dimensional lattice vector $\vec{\ell}$,
with graviton masses $m=|\vec \ell|/r_c.$ The first and second
$T_{\mu\nu}$'s are the stress tensors for the incoming $q\bar q,\ gg$
and outgoing $e^+e^-, \gamma\gamma$ states, respectively, and
$\sqrt{\hat s}$ is the parton center-of-mass energy. The sum on
$\vec{\ell}$ may be approximated by a continuous integration over KK
masses, modified by the cutoff, with the
result~\cite{Giudice:1998ck,Han:1998sg}
\begin{equation}
{\cal S}(\hat s)= \frac{S_{n-1}}{\md^{\,2+n}}\,
\int_0^{\infty}\frac{m^{n-1}\ dm\
e^{-m^2/\Lambda^2}} {m^2 - \hat  s} \ ,
\label{as1}
\end{equation}
where explicit integration over the $n-1$ angular variables leads to
the factor $S_{n-1} = 2\pi^{n/2} /\Gamma(n/2).$ The connection to an
effective four-point contact interaction in
Refs.~\cite{Giudice:1998ck,Han:1998sg, Hewett:1998sn} is made by
setting $\hat s=0$ in \eqref{as1}. This allows an explicit evaluation
of the integration over $m,$ with the result (for $n\ge 3$)
\begin{eqnarray}
{\cal S}(0) &=& \pi^{n/2}\
\frac{2}{n-2}\ \left(\frac{\Lambda}{\md}
\right)^{n-2}\ \frac{1}{\md^{\;4}}\nonumber\\
&\equiv& \frac{4\pi}{\Lambda_{T}^{\;4}}\ .
\label{amp}
\end{eqnarray}
In the last line we have used the convention of
Ref.~\cite{Giudice:1998ck} to parameterize the four-point
amplitude. At 95\% CL, the Tevatron data require $\Lambda_T>
1.2~\tev$. With the use of \eqref{amp}, this allows us to generate
Table~\ref{table1}, which shows the bounds on $\md$ for $n=3,\ldots,
7$ and $0.5 \le \Lambda/\md \le 1$. It is important to note that
(except for small variations for the case of
Ref.~\cite{Hewett:1998sn}, which permits a sign ambiguity in the
amplitude) {\em Table~\ref{table1} is independent of the conventions
in~\cite{Giudice:1998ck,Han:1998sg, Hewett:1998sn}.}  We can see that
the lower bounds on $\md$ depend on both $n$ and $\Lambda$. Typically,
$M_{D,{\rm min}} \alt 1~\tev$.

\begin{table}[tb]
\vspace*{-0.1in}
\begin{center}
\caption{95\% CL lower limits on $\md$ from the D\O\ Collaboration at
the Tevatron. }
\label{table1}
\vspace*{.1in}
\begin{tabular}{c|c|c|c|c|c}\hline\hline
\multicolumn{1}{c|}{$\quad\Lambda/\md\quad$}&
\multicolumn{5}{|c}{$M_{D,{\rm min}}$ (TeV)}
\\ \hline
& $\quad n=3 \quad$ & $\quad n=4 \quad$ & $\quad n=5\quad$
& $\quad n=6 \quad$ & $\quad n=7 \quad$ \\ \hline
0.5 & 0.98 & 0.80 & 0.70 & 0.63 & 0.58\\
0.6 & 1.02 & 0.88 & 0.80 & 0.76 & 0.73\\
0.7 & 1.06 & 0.95 & 0.90 & 0.89 & 0.88\\
0.8 & 1.10 & 1.01 & 1.00 & 1.01 & 1.04\\
0.9 & 1.13 & 1.07 & 1.09 & 1.14 & 1.21\\
1.0 & 1.16 & 1.13 & 1.18 & 1.26 & 1.38\\ \hline \hline
\end{tabular}
\end{center}
\end{table}

\section{BH production in particle collisions}
\label{sec:BH}

The preceding section discussed some potentially observable
consequences of scenarios with TeV-scale gravity, and the limits on
the scale of higher-dimensional gravity resulting from their
non-observation. In particular, for $n=4$ to 7, the quoted lower limit
on $\md$ comes from the non-observation at D\O\ of processes involving
KK gravitons. The spectrum and interactions of KK gravitons are
model-dependent, to an increasing degree at increasing scales above
$\md$.  Here we describe a more universal and model-independent
prediction of low-scale gravity scenarios: the production in particle
collisions of microscopic BHs.

\subsection{Geometric cross section}
\label{sec:sigma}

It has been argued~\cite{Banks:1999gd} that BH formation should occur
in the scattering of two incident particles when their impact
parameter is approximately less than the Schwarzschild radius of a BH
of mass equal to their center-of-mass energy $\sqrt{\hat{s}}$.  This
suggests a geometric cross section
\begin{equation}
\label{sigma}
\hat{\sigma} \approx \pi r_s^2 \ ,
\end{equation}
where
\begin{equation}
\label{schwarz}
r_s(\mbh) =
\frac{1}{\md}
\left[ \frac{\mbh}{\md} \right]^{\frac{1}{1+n}}
\left[ \frac{2^n \pi^{(n-3)/2}\Gamma({n+3\over 2})}{n+2}
\right]^{\frac{1}{1+n}}
\end{equation}
is the radius of a Schwarzschild BH of mass
$\mbh=\sqrt{\hat{s}}$~\cite{Myers:un,Argyres:1998qn} in $4{+}n$
dimensions.  Even if the incident particles are stuck on the SM brane,
the black hole formed should be treated as a fully $4{+}n$ dimensional
object in an asymptotically Minkowskian spacetime, as long as $r_s$ is
small compared to $r_c$.

The cross section~\eqref{sigma} grows like $\hat{s}^{1/(n+1)}$, more
rapidly than any SM cross section. Thus, at energies sufficiently far
above $\md$, BH production is expected to become the dominant
process. In $pp$ collisions at the LHC, rates as high as $10^8$ events
per year have been predicted in scenarios with $\md \sim
1~\tev$~\cite{Giddings:2001bu,Dimopoulos:2001hw}.

In our investigation of BH production by cosmic rays, we will be most
interested in collisions of neutrinos with atmospheric nucleons.
Since, at the energy scale of interest, gravitational cross sections
will be far smaller than the geometric area of a parton, we write the
$\nu N$ cross section as~\cite{Feng:2001ib}
\begin{equation}
\label{partonsigma}
\sigma ( \nu N \to \text{BH}) = \sum_i \int_{(\mbhmin{})^2/s}^1 dx\,
\hat{\sigma}_i ( \sqrt{xs} ) \, f_i (x, Q) \ ,
\end{equation}
where $s = 2 m_N E_{\nu}$, the sum is over all partons in the nucleon,
and the $f_i$ are parton distribution functions.  We set the momentum
transfer $Q = \min \{ \mbh, 10~\tev \}$, where the upper limit is from
the CTEQ5M1 distribution functions~\cite{Lai:2000wy}.  The cross
section $\sigma ( \nu N \to \text{BH})$ is highly insensitive to the
details of this choice~\cite{Feng:2001ib}. For example, choosing
instead $Q = \min \{ r_s^{-1}, 10~\tev \}$~\cite{Emparan:2001kf}
changes BH production rates by only 10\% to 20\%. For the conservative
$\nu$ fluxes considered below, our results are also insensitive to low
$x$. (For concreteness, however, we extrapolate to $x < 10^{-5}$
assuming a power law behavior $f_i(x,Q) \propto
x^{-[1+\lambda_i(Q)]}$.)  Finally, $\mbhmin$ is the minimal BH mass
for which \eqref{sigma} is expected to be valid.  The appropriate
choice of $\mbhmin$ is subject to theoretical uncertainties, as
discussed below.  We define
\begin{equation}
\xmin \equiv \mbhmin/\md
\end{equation}
and present results for various $1 \le \xmin \le 5$.  The dependence
of our event rates on $\xmin$ is found to be rather mild.

\begin{figure}[tbp]
\begin{minipage}[t]{0.49\textwidth}
\postscript{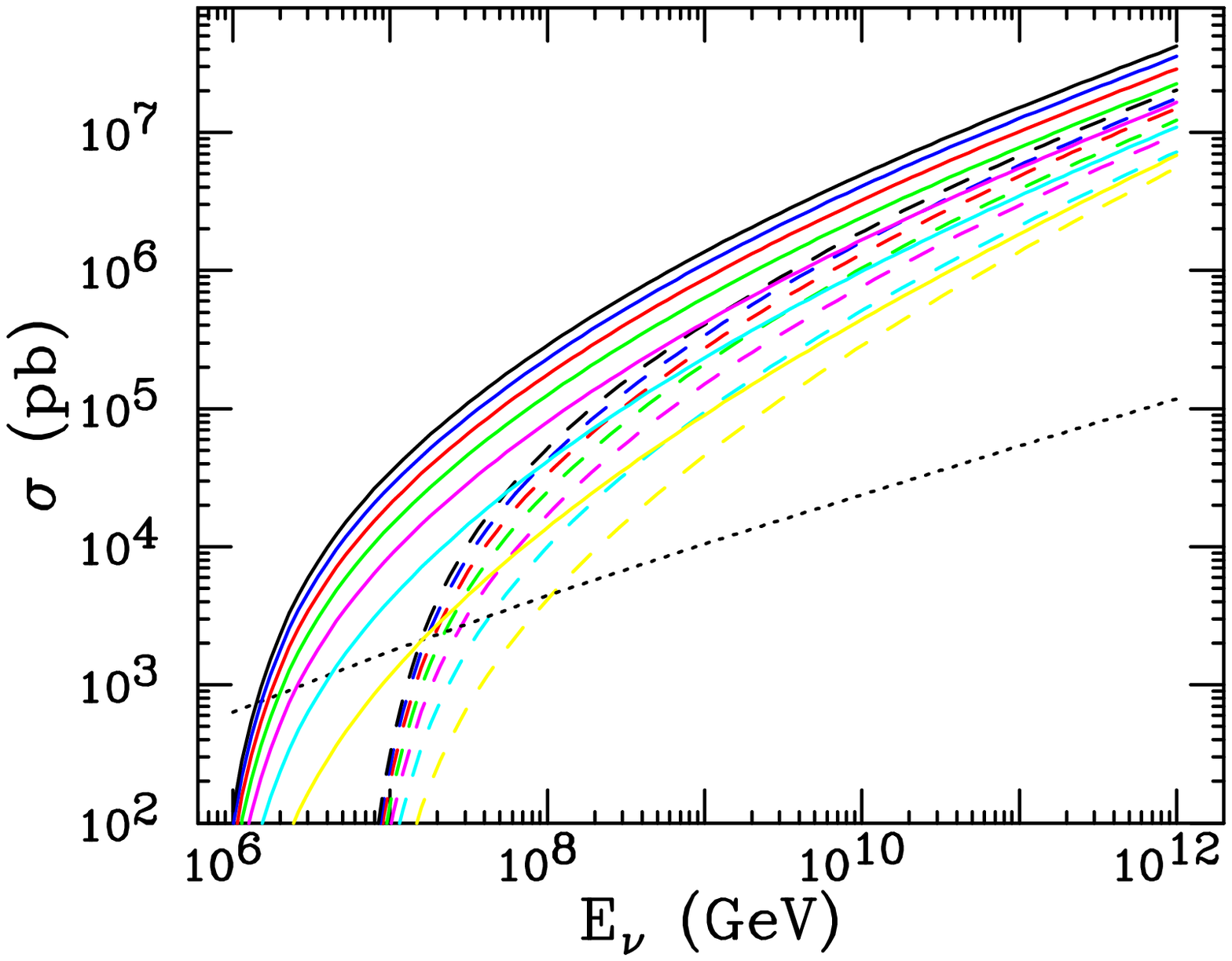}{0.99}
\end{minipage}
\hfill
\begin{minipage}[t]{0.49\textwidth}
\postscript{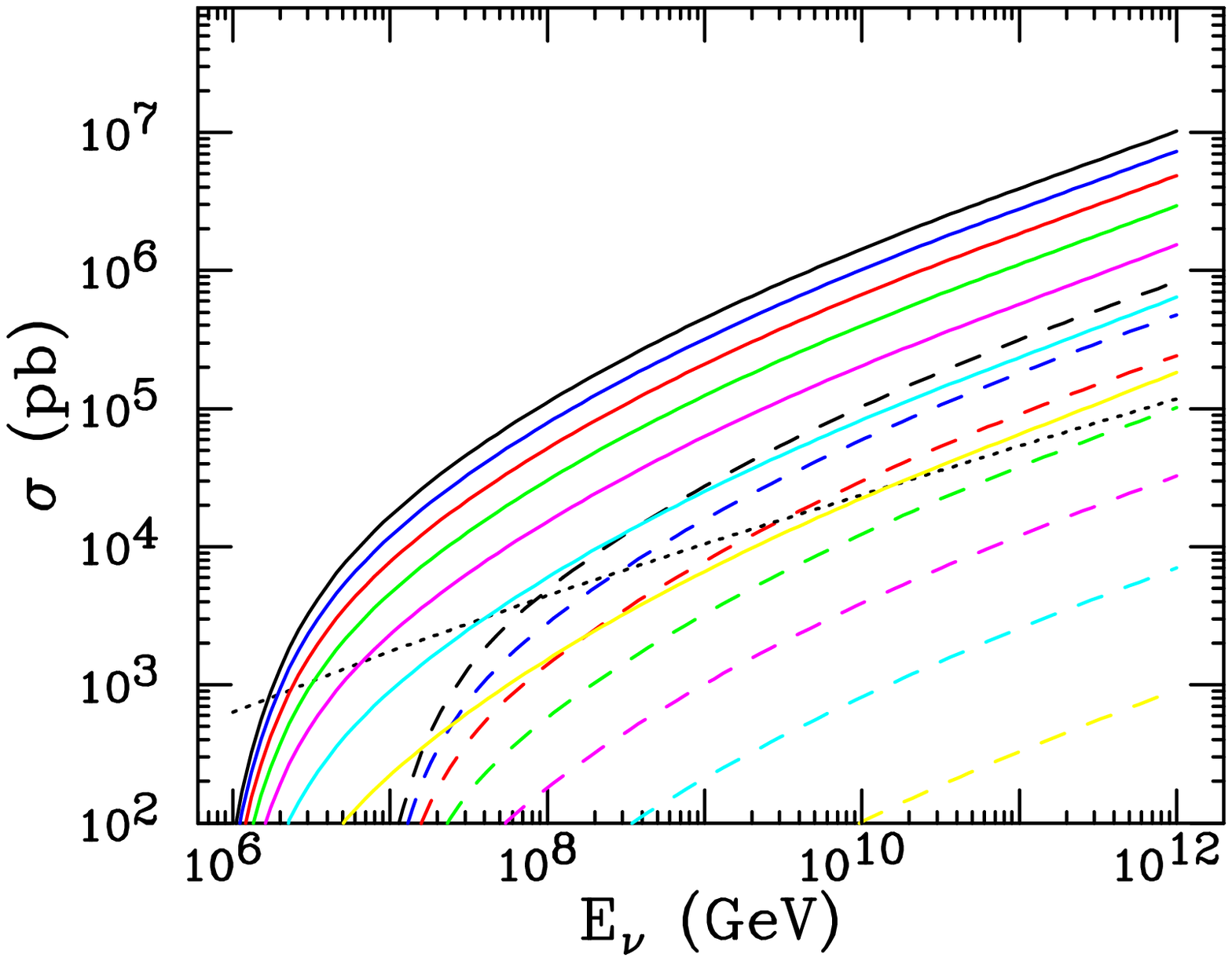}{0.99}
\end{minipage}
\caption{Cross sections $\sigma(\nu N \to {\rm BH})$ for $n=1,\ldots,
7$ from below for $\md = 1~\tev$, $\xmin=1$ (solid) and 3 (dashed),
and parton cross sections $\pi r_s^2$ (left) and $\pi r_s^2 e^{-I_E}$
(right). The SM cross section $\sigma(\nu N \to \ell X)$ (dotted) is
also shown.}
\label{fig:sigma}
\end{figure}

Cross sections for BH production by cosmic neutrinos are given in
Fig.~\ref{fig:sigma} for $\md = 1~\tev$; they scale as
\begin{equation}
\sigma(\nu N \to {\rm BH}) \propto \left[ \frac{1}{\md^2} \right]
^{\frac{2+n}{1+n}} \ .
\label{scaling}
\end{equation}
The SM cross section for $\nu N \to \ell X$ is also included for
comparison.  (Note that cross sections rise with increasing $n$ for
fixed $\md$, whereas they decrease for increasing $n$ for fixed
$\mstar$.)  As noted in Ref.~\cite{Feng:2001ib}, in contrast to the SM
process, BH production is not suppressed by perturbative couplings and
is enhanced by the sum over all partons, particularly the gluon.  As a
result of these effects, BH production may exceed deep inelastic
scattering rates in the SM by two or more orders of magnitude.

Although greatly reduced by the cross section for BH production,
neutrino interaction lengths
\begin{equation}
L=1.7\times 10^7~\kmwe~\left({\pb\over\sigma}\right)
\end{equation}
are still far larger than the Earth's atmospheric depth, which is only
$0.36~\kmwe$ even when traversed horizontally.  Neutrinos therefore
produce BHs uniformly at all atmospheric depths.  As a result, the
most promising signal of BH creation by cosmic rays will be
quasi-horizontal showers initiated by neutrinos deep in the
atmosphere.  For showers with large enough zenith angles, the
likelihood of interaction is maximized and the background from
hadronic cosmic rays is eliminated, since these shower high in the
atmosphere.

Once produced, the BH will Hawking evaporate, provided the
semiclassical approximation is valid.  In this case, a Schwarzschild
BH will behave like a thermodynamic system with temperature
\begin{equation}
T_H={n+1\over 4 \pi r_s}
\end{equation}
and entropy
\begin{equation}
\label{entropy}
S={2\pi^{(n+3)/2}\over 4 G_D\Gamma({n+3\over 2})} \,r_s^{n+2}
= {4\pi \, \mbh\ r_s\over n+2} \ .
\end{equation}

According to the semiclassical description, a BH produced in a
scattering event should be regarded as an intermediate state, which
decays on a time scale
\begin{equation}
\tau \sim {1\over \md} \left({\mbh\over \md}\right)^{3+n \over 1+n}\ .
\end{equation}
Since $\tau < 10^{-25}~\s$ for $\md \agt 1~\tev$ and $\mbh\alt
10~\tev$, the decay is effectively instantaneous.  In the decay
process, particles will be radiated into all available SM channels,
into quanta with energies typically of order $T_H$ or above.  These
decays are predicted to lead to highly distinctive signals in collider
events~\cite{Giddings:2001bu,Dimopoulos:2001hw,Cheung:2001ue}, with
high multiplicity, large transverse energy, hard leptons and jets, and
a characteristic ratio of hadronic to leptonic activity.

The magnitude of the entropy determines the validity of this picture.
Thermal fluctuations due to particle emission are small when $S \gg
1$~\cite{Preskill:1991tb}, and statistical fluctuations in the
microcanonical ensemble are small for $\sqrt S \gg 1$
\cite{Giddings:2001bu}. For $M_{BH}/\md=5$, \eqref{entropy} gives $S$
ranging from 29 for $n=4$ to 25 for $n=7$. For $M_{BH}/\md=3$ (or 1),
$S$ is about 13 (or 3) for $n$ between 4 and 7.

In searches for BH mediated events at colliders, it is essential to
set $\xmin$ high enough that the decay branching ratios predicted by
the semiclassical picture of BH evaporation are reliable, as there are
very large QCD backgrounds, and the extraction of signal from
background relies on knowing the BH decay branching ratios reliably.
This is especially true if one is attempting to determine discovery
limits, where the overall rates for BH production are not necessarily
large.  Thus, in collider searches, a cutoff of $\xmin = 5$ or more
may be appropriate.

By contrast, the search for deeply penetrating quasi-horizontal
showers initiated by BH decays can afford to be much less concerned
with the details of the final state, since the background is, relative
to colliders, almost nonexistent.  As a result, the signal relies only
on the existence of visible decay products, which, in this context,
includes all particles other than neutrinos, muons, and
gravitons. Indeed, there is very little about the final state, other
than its total energy and to some degree its multiplicity and
electromagnetic component~\cite{Anchordoqui:2001ei}, that we can
reasonably expect to observe, since detailed reconstruction of prompt
decay particles is not possible at cosmic ray detectors.  Thus, it
seems reasonable to choose a significantly lower value of $\mbhmin$
than is needed for collider searches; in our estimates of rates below
we will take $\xmin$ as low as 1.  While BHs of mass around $\md$ will
be outside the semiclassical regime, it seems quite reasonable to
expect that they, or their stringy progenitors, will nevertheless
decay visibly, whatever stringy or quantum gravitational description
applies.

As an illustration,  we examine the  scattering in the string
regime. For  $n=6$ large extra dimensions,  $M_s\sim
g_s^{1/4}M_{\rm Pl}$ $(M_s=$ string scale, $g_s=$ string
coupling). As shown in~\cite{Amati:1987wq}, the string cross
section $\sigma$ saturates to $\sim 1/M_s^2$ for $\sqrt{\hat s} >
M_s/g_s,$ (or in terms of $M_{\rm Pl},$ $\sigma \sim M_{\rm
Pl}^{-2}\,g_s^{-1/2}$ for $\sqrt{\hat s} > M_{\rm Pl}/g_s^{3/4}.$
As noted in~\cite{Dimopoulos:2001qe} this matches onto the
classical black hole cross section at an energy $M_s / g_s^2 \sim
M_{\rm Pl}/g_s^{7/4}.$ Thus, if $g_s$ is not too small (implying
a small hierarchy between $M_s$ and $M_{\rm Pl}$),  the
transition to the geometric cross section is rapid. In this work
we adopt a minimum energy $\sqrt{\hat s}\sim 1-3\ M_D\simeq 3-9
M_{\rm Pl}$ (for $n=6),$  so that while there are probably stringy
corrections, the cross section should be substantially geometric.

\subsection{Uncertainties in the Cross Section}
\label{sec:uncertainties}

Although the details of the process by which BHs decay are not of
great concern to us, the production process is of central importance,
since our rates (and the lower limits we will be able to set on $\md$)
will depend directly on the form of the cross section.  It should be
emphasized that the heuristic arguments on which \eqref{sigma} is
based only determine $\sigma$ up to an overall factor of order
one. Even at the classical level, our conclusions could be
significantly affected by theoretical uncertainties in this factor,
four sources of which we now discuss.

\subsubsection{Mass ejection}

Classical general relativity calculations~\cite{D'Eath:hb} indicate
that the mass of a BH formed in a head-on collision is somewhat less
(about 16\% less) than the total center-of-mass energy. At least in
four dimensions, this suggests that the formula \eqref{sigma} should
be modified by replacing $r_s(\mbh)$ by $r_s(0.84 \mbh)$, leading to a
slight reduction of $\sigma$.  Very recently, corresponding
calculations in more than four dimensions have been
presented~\cite{Eardley:2002re}.

\subsubsection{Angular momentum}

The analytic techniques used to study head-on collisions are not
applicable to collisions at nonzero impact parameter. Thus the claim
that a BH will be produced when $b\alt r_s(\sqrt{\hat{s}})$ can only
be expected to be true up to a numerical factor.

One issue that arises at nonzero impact parameter that we can address
is that the BHs formed will have angular momentum~\cite{Park:2001xc}.
In particular, the Schwarzschild radius appearing in the formula
\eqref{schwarz} should more accurately be replaced by the radius of a
Kerr BH of the appropriate angular momentum.  This will alter the
critical impact parameter at which a BH will form, for given
$\hat{s}$.  For two particles each of energy $E$ in the center-of-mass
frame colliding with impact parameter $b$, the total angular momentum
with respect to the center of mass is $J=2(b/2)E=b \mbh/2$.  So the
maximum impact parameter at which a BH will form should occur at a
value of $b$ for which the radius $r_k(M,J=b \mbh/2)$ of a Kerr BH and
$b$ are equal.  The Kerr radius satisfies~\cite{Myers:un}
\begin{equation}
\mbh = c_n r_k^{n-1} \left[ r_k^2 + (n+2)^2 J^2 / 4 \mbh^2 \right]
= c_n r_k^{n-1} \left[ r_k^2 + (n+2)^2 b^2 / 16 \right]
\end{equation}
with $c_n$ an $n$-dependent constant. Setting $r_k=b$ we get
\begin{equation}
\mbh = c_n r_k^{n+1} \left[ 1+ (n+2)^2  / 16 \right] \ .
\end{equation}
Since for a Schwarzschild BH,
$\mbh = c_n r_s^{n+1}$,
the cross section in \eqref{sigma} should be corrected to
\begin{equation}
\hat{\sigma} \approx \pi r_k^2(\mbh,J)=
\left[ 1+ (n+2)^2  / 16 \right]^{-{2\over n+1}}\, \pi r_s^2(\mbh) \ .
\end{equation}
For $1 \le n \le 7$, the correction factor is remarkably stable,
ranging from 0.62 to 0.64. This result has been recently confirmed
~\cite{Eardley:2002re} in a
classical analysis of black hole formation for collisions with non-zero
impact parameter.  We see, then, that including the effect
of angular momentum also leads to a small reduction of $\sigma$.

\subsubsection{Sub-relativistic limit}

While the corrections related to mass ejection and angular momentum
both appear to decrease $\sigma$ by factors of order 1, another
potential correction to the naive cross section \eqref{sigma} could
enhance it by a more-than-compensating factor. Namely, the critical
impact parameter may be somewhat larger than the radius of the BH
formed.  An argument that supports this conjecture may be given in the
case where the incident particles are sub-relativistic with rest mass
approximately $E=\mbh/2$. In this case, the incident particles may be
treated as BHs with mass $\mbh/2$. If $b\le 2r_{s}(\mbh/2)$, they will
touch as they pass, and thus merge into a BH of mass $\mbh$. In this
regime, we expect as an approximate lower bound on $\sigma$ of
\begin{equation}
\sigma \agt \pi b^2 = \pi [2r_s(\mbh/2)]^2= 4^{n/(n+1)}\pi
r_s(\mbh)^2 \ .
\end{equation}
For large $n$, the correction factor approaches 4. Of course, the
situation may change considerably in the ultra-relativistic limit, but
this estimate, in a limit that we understand, at least makes it
plausible that the correct coefficient in \eqref{sigma} might be
larger than 1.  Thus our choice to take $\sigma=\pi r_s^2$ may well
turn out to be conservative.

\subsubsection{Gravitational infall and capture}

It could also be argued that the cross section of \eqref{sigma} is too
small, because it supposes that a black hole only forms when the two
particles come within a Schwarzschild radius of each other, when in
fact we expect that gravitational collapse will occur for somewhat
larger impact parameters as well. Another problem with \eqref{sigma}
is that while the cross section is measured with respect to the flat
geometry of the asymptotic region, the Schwarzschild radius is a
property of the highly curved region close to the
singularity~\cite{Emparan}.

A better measure of the cross-sectional area associated with a black
hole of given mass, which overcomes these objections, is given by the
classical cross section for photon capture~\cite{Wald:rg}.  If a beam
of parallel light rays is sent in towards a Schwarzschild black hole
from the asymptotically Minkowskian region of spacetime, the black
hole's classical cross section is defined to be the cross-sectional
area of the portion of the beam that gets captured.  In four
dimensions, one finds from the geodesic equation (see, \eg,
Ref.~\cite{Wald:rg}) that
\begin{equation}
\sigma =\pi b_c^2 = 27 \pi G_4^2 \mbh^2 \ ,
\end{equation}
independent of the energies of the incoming photons. (The
relevance of this cross section for black hole production has
been independently argued in Ref.~\cite{Solodukhin:2002ui}.) The
maximum impact parameter $b_c$ at which capture occurs is about
2.6 times as large as the Schwarzschild radius, and the cross
section is enhanced by a factor of $27/4$.  A straightforward
extension of this calculation to Schwarzschild black holes in
$4+n$ dimensions gives~\cite{Emparan:2000rs}
\begin{equation}
b_c
= {\frac
{(3+n)^{\frac{1}{2}\left({3+n\over 1+n}\right)}}
{2^{\frac{1}{1+n}}\sqrt{1+n}}}
\, r_s \ .
\end{equation}
For $n=1$, the cross section is 4 times larger than the cross section
of \eqref{sigma}, and for $n=7$, it is still 87\% larger.  Thus, the
enhancement suggested by this definition of the cross section could be
significant and could easily offset other possible
reductions.\footnote{A somewhat more refined estimate of the cross
section would take into account the rotation of the black hole.  One
may derive a geodesic equation for null geodesics in the equatorial
plane of a rotating black hole (since the incoming particles {\em are}
in the equatorial plane), and calculate the impact parameter
$b(\mbh,a)$ at infinite distance from a black hole of mass $\mbh$ and
rotation parameter $a$ (as defined in
Ref.~\cite{Wald:rg,Myers:un}). In four dimensions, the extremal value
$a=\mbh$ gives $b=2\mbh$, reproducing the cross section of
\eqref{sigma}.}

\subsection{Exponential suppression?}
\label{sec:Voloshin}

Quantum mechanical corrections to the amplitude for BH production may
be even more significant than classical uncertainties.  In particular,
Voloshin~\cite{Voloshin:2001vs,Voloshin:2001fe} has proposed that the
cross section of \eqref{sigma} should be modified by an exponential
suppression factor
\begin{equation}
\label{sigmasupp}
\sigma \sim \pi r_s^2 e^{-I_E} \ ,
\end{equation}
with $I_E$ the Euclidean Gibbons--Hawking action for the BH,
which, in terms of the entropy of \eqref{entropy}, is
\begin{equation}
I_E={S\over n+1}\ .
\end{equation}

In part, Voloshin's critique is based on previous attempts to
calculate amplitudes for the production of classical field
configurations (in which the multiplicity is greater than the inverse
coupling) from initial quantum states. The intrinsically
nonperturbative nature of such processes suggests employing an
instanton--like approximation, which could lead to exponential
suppression.  Such an approach can be taken even for processes that
are semi-classically allowed (such as multi-Higgs production
\cite{Cornwall:1990hh,Goldberg:1990qk}) that do not require tunneling
in order to take place.  The problem is not yet solved. For example, a
recent lattice simulation~\cite{Charng:2001ht} shows no evidence for
the enhancement of large multiplicity amplitudes manifest in
perturbation theory, perhaps counter-indicating the formation of a
classical field state in the quantum collision.

We are cognizant of the uniqueness of gravitation (such as the onset
of strong coupling for $\hat s> M_D^2$), and the support in favor of
the geometric cross section based on classical calculations for both
vanishing~\cite{D'Eath:hb} and non-vanishing~\cite{Eardley:2002re}
impact parameter. An additional supporting argument based on a string
calculation has been given~\cite{Dimopoulos:2001qe}, and the
applicability of CPT arguments when comparing black hole formation and
decay, an element in Voloshin's criticism~\cite{Voloshin:2001vs}, has
been questioned~\cite{Giddings:2001ih,Solodukhin:2002ui}.
Nevertheless, for completeness, we will also present results below for
the exponentially suppressed cross section of \eqref{sigmasupp}. For
cosmic rays, we will see that even with Voloshin's suppression factor
included, useful bounds will emerge after 5 years of operation of
Auger.

\section{The cosmogenic neutrino flux}
\label{sec:flux}

Among the many possible sources of ultra-high energy neutrinos, the
cosmogenic flux is the most reliable.  This neutrino flux relies only
on the assumption that the observed extremely high energy cosmic rays
contain nucleons and are primarily extragalactic in origin. If the
charge of the primaries satisfies $Z \alt {\cal O}(1)$, as recently
reported~\cite{Ave:2001sd}, an extragalactic origin is almost
guaranteed, as the observed nearly isotropic angular distribution
strongly disfavors galactic disk sources~\cite{Takeda:1999sg}.
Moreover, even if the absence of the Greisen-Zatsepin-Kuzmin (GZK)
cutoff~\cite{Greisen:1966jv} on cosmic ray energies is a reflection of
our coincidental position near a nucleus/nucleon-emitting source, one
still expects the full cosmogenic neutrino flux.

Briefly, the argument for this is as follows~\cite{Ahn:1999jd}: the
known astrophysical environments (within a few Mpc of the Earth) are
not among the most powerful, but (in principle) can produce hadronic
cosmic rays with the desired energies when parameters are stretched to
their limits.  Thus if these less powerful sources can accelerate
particles above $10^{20}~\ev$, it must be that more powerful distant
sources (like Fanaroff-Riley II radiogalaxies) can accelerate protons
above photopion threshold, giving rise to the cosmogenic neutrino
flux.  Moreover, the approximately smooth power law behavior of the
observed spectrum above $10^{19}~\ev$~\cite{Nagano:ve} seems to
indicate that any ``local source'' contribution should be comparable
to that of all other sources in the universe.  Otherwise, one should
invoke an apparently miraculous matching of spectra to account for the
smoothness of the spectrum. This smoothness will provide the basis for
obtaining the cosmogenic neutrino flux, as discussed in what follows.

The chain reaction generating these cosmogenic neutrinos, triggered by
GZK pion photo-production, is well
known~\cite{beresinsky,Stecker:1979ah}.  The resulting neutrino flux
depends critically on the cosmological evolution of the cosmic ray
sources and on their proton injection
spectra~\cite{Yoshida:pt,Protheroe:1996ft,Engel:2001hd}. The high
energy tail of the neutrino spectrum can also receive a significant
contribution from semi-local sources, such as the Virgo
cluster~\cite{Hill:1985mk}.  Additionally, there is a weak dependence
on the details of the cosmological expansion of the universe.  For
example, a small cosmological constant tends to increase the
contribution to neutrino fluxes from higher
redshifts~\cite{Engel:2001hd}.

\begin{figure}[tbp]
\postscript{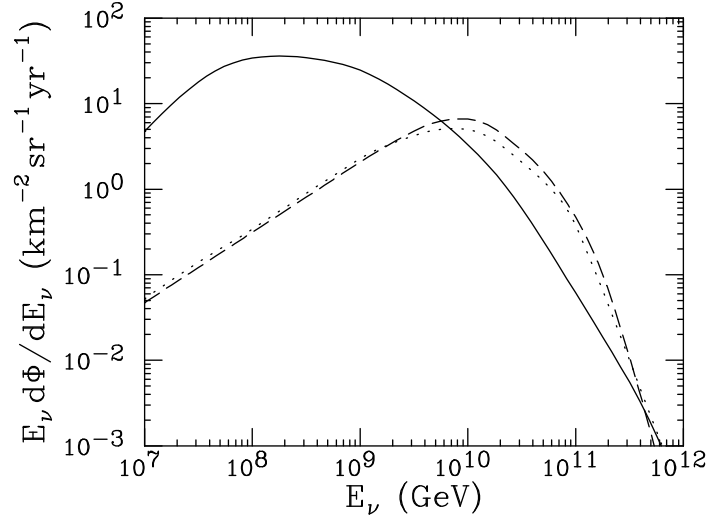}{0.56}
\caption{Cosmogenic $\nu_{\mu} + \bar{\nu}_{\mu} + \nu_e$ fluxes from
Protheroe and Johnson with energy cutoff of $3 \times 10^{21}~\ev$
(solid)~\protect\cite{Protheroe:1996ft}, Hill and Schramm
(dashed)~\protect\cite{Hill:1985mk}, and previous estimate by Stecker
without source evolution (dotted)~\protect\cite{Stecker:1979ah}.  See
text for discussion. }
\label{fig:flux_bhauger}
\end{figure}

In our analysis we adopt the cosmogenic neutrino flux estimates of
Protheroe and Johnson (PJ)~\cite{Protheroe:1996ft}.  We consider their
$\nu_{\mu} + \bar{\nu}_{\mu}$ estimate with an injection spectrum with
$E_{\rm cutoff} = 3 \times 10^{21}~\ev$.  In addition to $\nu_\mu$ and
$\bar{\nu}_\mu$, electron neutrinos also contribute to black hole
production. In the high energy peak, the $\nu_e$, $\nu_\mu$, and
$\bar{\nu}_\mu$ fluxes are nearly identical~\cite{Engel:2001hd}, and
we include this $\nu_e$ flux in our analysis.  The study of PJ
incorporates the source cosmological evolution from
estimates~\cite{Rachen:1992pg} of the power per comoving volume
injected in protons by powerful radio galaxies, taking into account
the radio luminosity functions given in Ref.~\cite{Peacock}.  The
shape of the resulting neutrino spectrum is shown in
Fig.~\ref{fig:flux_bhauger}. The flux peaks around $E \approx 2 \times
10^{17}$ eV, which is roughly the same energy suggested by other
analyses following a source evolution proportional to
$(1+z)^4$~\cite{Yoshida:pt,Engel:2001hd}.  To explore possible
additional contributions from semi-local nucleon sources, we also
consider below the cosmogenic neutrino flux estimates of Hill and
Schramm (HS)~\cite{Hill:1985mk}, which are also given in
Fig.~\ref{fig:flux_bhauger}.  A flux estimate of Stecker is also given
there.  As noted in~\cite{Feng:2001ib}, the PJ, HS, and Stecker fluxes
all yield approximately the same rates for BH production.

We stress that the PJ flux agrees with the most recent
estimate~\cite{Engel:2001hd} in the entire energy range, whereas the
spectrum obtained in earlier calculations~\cite{Yoshida:pt} is
somewhat narrower, probably as a result of different assumptions
regarding the propagation of protons. The PJ analysis is performed
within Friedmann cosmology with vanishing cosmological constant
$\Lambda$, $q_0 =0.5$, and $H_0 = 75~\km~\s^{-1}$ Mpc$^{-1}$, assuming
an extragalactic magnetic field of 1 nG and a source spectrum
proportional to $E^{-2}$ up to redshifts $z=9$. The extension to
cosmological models with $\Lambda \neq 0$ would not produce remarkable
changes. For example, for $\Omega_{\rm M} =0.3$ and $\Omega_\Lambda =
0.7$, the neutrino flux is increased by a factor of $<
1.7$~\cite{Engel:2001hd}.

\section{Acceptance of surface arrays for neutrino showers}
\label{sec:acceptance}

Ultra-high energy cosmic neutrinos may be detected by ground arrays
and fluorescence detectors on the surface of the Earth, as well as by
space-based fluorescence detectors, and neutrino telescopes beneath
the Earth's surface.  Here we concentrate on ground arrays, and
consider two prominent examples: AGASA, the largest surface array
currently in operation, and the Auger Observatory now under
construction.

AGASA consists of 111 scintillation detectors each of area 2.2 m$^2$,
spread over an area of $100~\km^2$ with 1 km
spacing~\cite{Chiba:1991nf}.  The array detectors are connected and
controlled through a sophisticated optical fiber network. The array
also contains a number of shielded scintillation detectors which
provide information about the muon content of the showers. The full
AGASA experiment has been running since 1992, and has recorded the
majority of events claimed to have energies above the GZK cutoff.

The Auger Observatory is a hybrid experiment, with two sites (one in
the northern hemisphere and one in the southern), each covering an
area of $3000~\km^2$ and consisting of 1600 particle detectors
overviewed by 4 fluorescence detectors~\cite{auger}.  The surface
array stations are cylindrical water \v{C}erenkov detectors with area
10 m$^2$, spaced 1.5 km from each other in an hexagonal grid.  Event
timing is made possible through global positioning system
receivers. The optical system uses the fluorescence technique
pioneered by the University of Utah's Fly's Eye
detector~\cite{Baltrusaitis:mx}. ``Golden events,'' events detected by
both methods simultaneously, will be extremely valuable for
experimental calibration.  However, atmospheric fluorescence detection
is possible only on clear, dark nights, and so the golden event rate
is expected to be less than 10\% of the total event rate. We consider
only the ground array below.  The full southern site is scheduled for
completion in 2003. Its engineering array, at 1/40 of the full size,
is now complete and is already detecting giant air
showers~\cite{Allekote}.

\begin{figure}[tbp]
\postscript{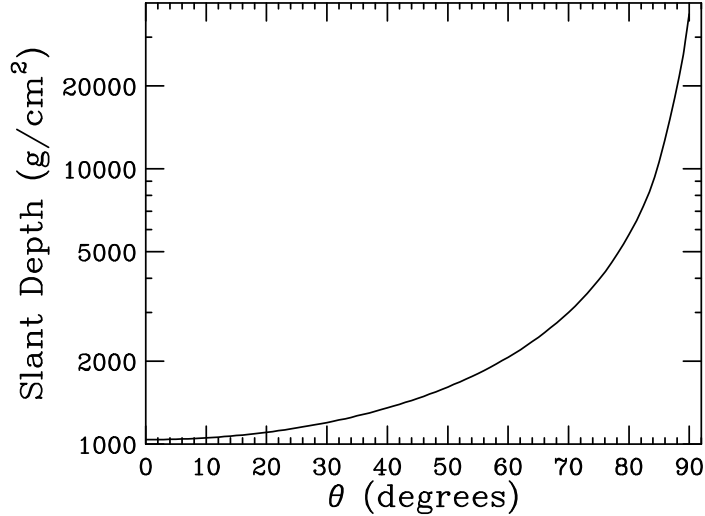}{0.56}
\caption{Slant depths corresponding to various zenith angles
$\theta$. }
\label{fig:slantdepth}
\end{figure}

A surface array's acceptance for neutrino detection may be expressed,
in units of $\km^3$ water equivalent steradians ($\km^3{\rm we}~\sr$),
as~\cite{Billoir:nq}
\begin{equation}
A(E) =  S
\int_{\theta_{\rm min}}^{\theta_{\rm max}} 2\pi \sin\theta \, d\theta
\int_0^{h_{\rm max}} \frac{\rho_0}{\rho_{\rm water}}
\, e^{-h/H}\, {\cal P} (E,\theta,h)\, dh \ ,
\label{a}
\end{equation}
where $S$ is the area of the ground array, $\rho_0 \approx 1.15 \times
10^{-3} \rho_{\rm water}$ is the density of the atmosphere at ground
level, $H \approx 8~\km$, $h_{\rm max} = 15~\km$, and ${\cal P}
(E,\theta,h)$ is the probability of detecting a shower with energy $E$
and zenith angle $\theta$ that begins at altitude $h$. The minimum
zenith angle is set by the desire to separate deep neutrino-initiated
showers from far showers initiated by hadronic primaries.  Typically,
a minimum zenith angle in the range $60^{\circ} < \theta_{\rm min} <
75^{\circ}$ is imposed. This range corresponds to atmospheric slant
depths of 2000 to 4000 $\g/\cm^2$.  (See Fig.~\ref{fig:slantdepth}.)
The maximum zenith angle $\theta_{\rm max}$ varies from analysis to
analysis.  For example, in an analysis of fully contained showers, a
value of $\theta_{\rm max}$ below $90^{\circ}$ is required for showers
to deposit all of their energy within the array.

\begin{figure}[tbp]
\postscript{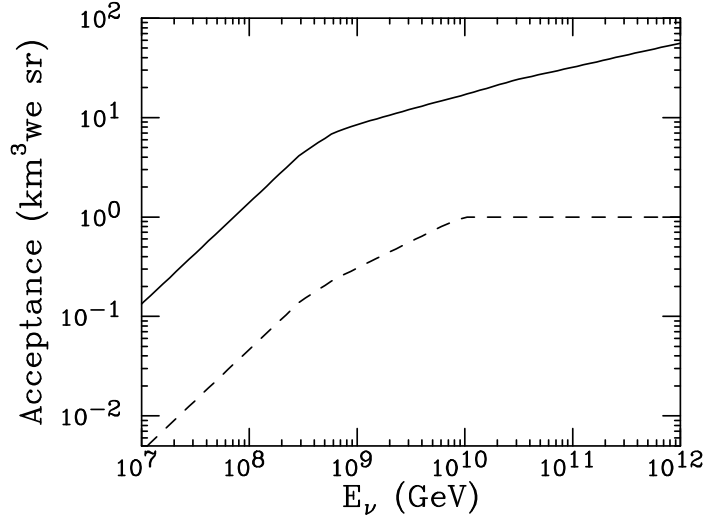}{0.56}
\caption{Ground array acceptances for quasi-horizontal air showers at
the Auger Observatory (solid) and AGASA (dashed).  See text for
discussion. }
\label{fig:accept_bhauger}
\end{figure}

Reliable Monte Carlo simulations to determine Auger's acceptance for
quasi-horizontal showers have been performed by several
groups~\cite{Billoir:nq,Capelle:1998zz}. Of course, the acceptance
depends on the amount and type of energy generated by neutrino
interactions in the atmosphere. For example, the charged current
interaction $\nu_\mu p \to \mu^+ X$ produces a muon that carries
approximately 80\% of the incoming energy and is not detectable at
Auger.  Acceptances for both electromagnetic and hadronic showers have
been determined in Ref.~\cite{Capelle:1998zz}. BHs decay thermally,
according to the number of degrees of freedom available, and so their
decays are mainly hadronic. We therefore adopt the hadronic acceptance
of Ref.~\cite{Capelle:1998zz} including partially contained showers
with zenith angles $\theta > 75^{\circ}$. The acceptance is given in
Fig.~\ref{fig:accept_bhauger}. Partially contained showers, where the
shower axis does not pass through the array, do not contribute
significantly to the Auger acceptance for shower energies below
$10^{10}~\gev$.

The AGASA Collaboration has searched for deeply penetrating
showers~\cite{Inoue:cn,agasa}.  In these studies, they find that, for
showers with energy above $10^{10}~\gev$ and the requisite zenith
angle, the detection probability ${\cal P}$ becomes effectively 100\%
and independent of altitude~\cite{agasa}. For these energies, then,
\eqref{a} may be re-written as
\begin{equation}
A(E > 10^{10}~\gev) \approx (S\Omega)_{\rm eff}(E)
\int_0^{h_{\rm max}} \frac{\rho_0}{\rho_{\rm water}} \,
e^{-h/H}\, dh \ ,
\label{b}
\end{equation}
where
\begin{equation}
(S\Omega)_{\rm eff} (E) \equiv S\,
\int_{\theta_{\rm min}}^{\theta_{\rm max}} 2\pi \sin\theta \,
{\cal P} (E,\theta)\, d\theta
\end{equation}
is the ``effective area $\times$ solid angle''~\cite{MC}.  Acceptances
for extremely energetic showers may then be quoted in terms of
$(S\Omega)_{\rm eff}$.  The AGASA Collaboration has searched for
deeply penetrating showers of any origin in Ref.~\cite{Inoue:cn}.
They find none with energy above $10^{10}~\gev$ in $9.7 \times
10^{7}~\s$ of exposure.  Given an upper bound of 2.44 events at 90\%
CL, then, they derive a flux limit for deeply penetrating showers of
$1.9 \times 10^{-10}~\km^{-2}~\sr^{-1}~\s^{-1}$, which implies
$(S\Omega)_{\rm eff} = 132~\km^2~\sr$ for quasi-horizontal air
showers. Equivalently, given \eqref{b}, the AGASA acceptance for
neutrino initiated events is
\begin{equation}
A(E>10^{10}~\gev) \approx 1.0~\km^3{\rm we}~\sr \ .
\label{c}
\end{equation}

This acceptance is roughly 30 times smaller than the neutrino
acceptance of Auger, as one would naively guess from the ratio between
the Auger and AGASA surface areas. For lower energies, since the
separation between detectors is smaller at AGASA than at Auger, a
conservative approach is to model the AGASA acceptance as that of
Auger reduced by a factor of 30.  We adopt this estimate for energies
below $5 \times 10^8~\gev$, and interpolate smoothly between this and
\eqref{c} for energies $5 \times 10^8~\gev < E_{\nu} < 10^{10}~\gev$.
The resulting AGASA acceptance is shown in
Fig.~\ref{fig:accept_bhauger}.

\section{New Bounds from AGASA}
\label{sec:AGASA}

Given the cross sections, apertures, and fluxes discussed above, the
number of BHs detected by a given experiment is
\begin{equation}
N = \int dE_{\nu}\, N_A \, \frac{d\Phi}{dE_{\nu}} \,
\sigma(E_{\nu}) \, A(E_{\nu}) \, T \ ,
\label{numevents}
\end{equation}
where $A(E_{\nu})$ is the experiment's acceptance in cm$^3$we sr, $N_A
= 6.022 \times 10^{23}$ is Avogadro's number, $d\Phi/dE_{\nu}$ is the
source flux of neutrinos, and $T$ is the running time of the detector.
We now determine current bounds on BH production from the
AGASA experiment.  In the next section, we examine future prospects
for BH detection at the Auger Observatory.

The AGASA Collaboration has searched for deeply penetrating
quasi-horizontal showers~\cite{agasa}.  The depth at which a shower is
initiated is, of course, not directly measurable in a ground array.
However, the electromagnetic components of far showers are
extinguished by ground level, leaving only a muon component, whereas
for deeply penetrating showers, both electromagnetic and muon
components are detected.  By exploiting this difference, deeply
penetrating quasi-horizontal showers may be distinguished from showers
induced by hadronic cosmic rays.

Relative to showers with muon components only, showers with
eletromagnetic components have charged particle densities that are
more concentrated near the shower axis, and their shower fronts are
more curved.  The depth at shower maximum $\xmax$ may then be
determined through its correlation to two measurable quantities:
$\eta$, which parameterizes the lateral distribution of charged
particles, and $\delta$, which parameterizes the curvature of the
shower front.  The values of $\xmax$ as determined by these
correlations, denoted $\xmax^{\eta}$ and $\xmax^{\delta}$, are then
required to be large to distinguish candidate neutrino events from
showers induced by hadronic cosmic rays.

In 1710.5 days of data recorded from December 1995 to November 2000,
the AGASA Collaboration found 6 candidate events with $\xmax^{\eta},
\xmax^{\delta} \ge 2500~\g/\cm^2$.  The expected background from
hadronic showers is $1.72{}^{+0.14}_{-0.07}{}^{+0.65}_{-0.41}$, where
the first uncertainty is from Monte Carlo statistics, and the second
is systematic.  Of the 6 candidate events, however, 5 have values of
$\xmax^{\eta}$ and/or $\xmax^{\delta}$ that barely exceed
$2500~\g/\cm^2$, and are well within $\Delta \xmax$ of this value,
where $\Delta \xmax$ is the estimated precision with which $\xmax$ can
be reconstructed.  The AGASA Collaboration thus concludes that there
is no significant enhancement of deeply penetrating shower rates given
the detector's resolution.

The AGASA results imply lower bounds on the scale of low-scale
gravity, assuming the conservative cosmogenic fluxes of
Sec.~\ref{sec:flux}.  For these fluxes, the expected rate for deeply
penetrating showers at AGASA from SM neutrino interactions is about
0.02 events per year, and so negligible.  Given 1 event that
unambiguously passes all cuts, and the central value of 1.72
background events, the AGASA results imply an upper bound of 3.5 black
hole events at 95\% CL~\cite{Feldman:1997qc}.

\begin{figure}[tbp]
\postscript{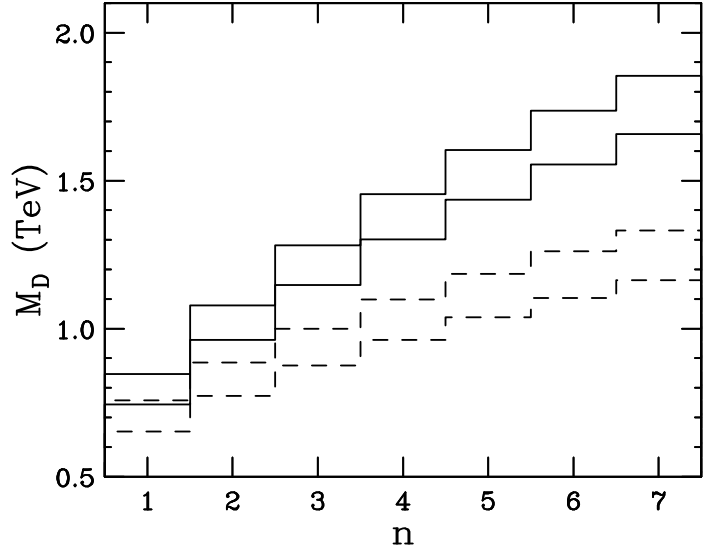}{0.56}
\caption{95\% CL lower bound on $\md$ from non-observation of
quasi-horizontal air showers in 1710.5 live days at AGASA for
$\xmin=1$ (solid) and 3 (dashed), assuming the cosmogenic neutrino
flux of Protheroe and Johnson (lower) and Hill and Schramm (upper).}
\label{fig:AGASA_n_MD}
\end{figure}

The 3.5 event contour is given for various dimensions $n$ in
Fig.~\ref{fig:AGASA_n_MD}.  For $\xmin = 1$, the absence of deeply
penetrating showers in the AGASA data implies
\begin{eqnarray}
n=4 &:& \quad \md > 1.3 - 1.5~\tev \nonumber\\
n=7 &:& \quad \md > 1.6 - 1.8~\tev \ .
\label{AGASAbounds}
\end{eqnarray}
Results for $\xmin =3$ are also given in Fig.~\ref{fig:AGASA_n_MD}.
They imply $\md > 1.0 - 1.1~\tev$ for $n = 4$, and $\md > 1.1 -
1.3~\tev$ for $n = 7$; even for $\xmin =3$, these bounds are exceed or
are competitive with all existing collider and astrophysical bounds.
As argued in Sec.~\ref{sec:sigma}, $\xmin = 1$ is a reasonable
assumption for the present application, as the derivation of limits
relies only on the assumption that BHs or their lighter progenitors
with mass around $\md$ decay visibly.  This assumption is violated
only if their decays are limited to neutrinos, gravitons, and muons.

The range in \eqref{AGASAbounds} is from considering both PJ and HS
fluxes.  As noted in Ref.~\cite{Feng:2001ib}, the dependence of the
bounds on variations in the evaluations of cosmogenic fluxes is
weak. These bounds are conservative in that larger non-cosmogenic
fluxes, as predicted by some models and as may be indicated by
super-GZK cosmic rays, will strengthen them, possibly dramatically.
Note also that we have neglected enhancements to cosmic neutrino
interactions from sub-Planckian extra-dimensional physics, which are
more model-dependent, but can only serve to strengthen these bounds.

The bounds of \eqref{AGASAbounds} are, of course, subject to the
${\cal O}(1)$ uncertainties inherent in the parton level cross
section.  Given this cross section, however, they are direct bounds on
the fundamental Planck scale $\md$, and are not subject to the
uncertainties inherent in collider bounds, such as the choice of brane
softening parameter $\Lambda$ discussed in Sec.~\ref{sec:tevatron}.
Any comparison of collider and cosmic ray
bounds is then subject to the independent
uncertainties associated with each bound.
Nevertheless, for $n \ge 4$, given the geometric BH cross section, the AGASA limit is more stringent that all existing collider bounds for all choices of
$\Lambda/\md \le 1$.

Before leaving the AGASA results, we derive their implications for
extra dimensions if taken at face value.  Given 6 events with an
expected background of 1.72 events, the expected signal is 0.86 to 11
events at 95\% CL.  The preferred region of the $(n, \md)$ plane is
given in Fig.~\ref{fig:AGASA_preferred}. (In
Fig.~\ref{fig:AGASA_preferred}, and all following figures, we use the
PJ flux.  The HS flux yields slightly larger rates.)  The evidence for
BH production (or any other anomaly) is speculative, given the
statistics and the peculiarities of the data noted above.  However,
this analysis shows the power of cosmic ray measurements for probing
extra dimensions.  The preferred Planck scales are not probed by any
other experiment.  At the same time, they will be thoroughly explored
in the near future at larger cosmic ray experiments, such as the Auger
Observatory, to which we now turn.

\begin{figure}[tbp]
\postscript{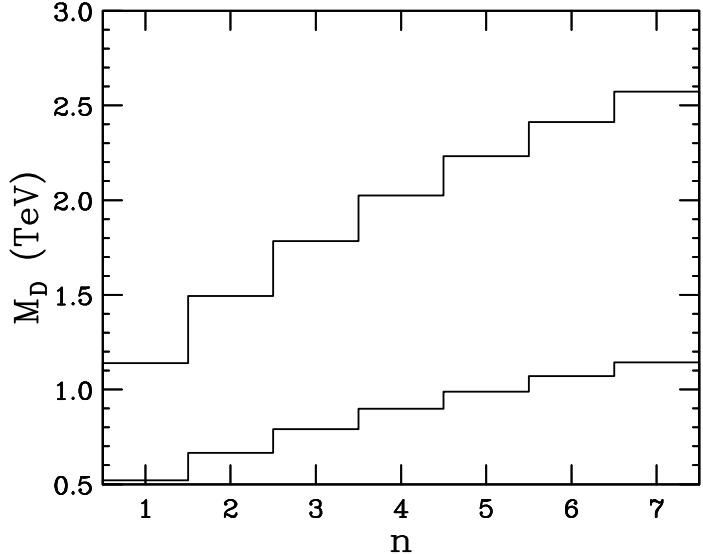}{0.56}
\caption{
\protect\label{fig:AGASA_preferred}
95\% CL upper and lower bounds on $\md$ for various $n$,
given 6 candidate events above a background of 1.72 in 1710.5 live
days at AGASA, and ascribing the excess to BH production.  We
assume $\xmin=1$ and the cosmogenic neutrino flux of Protheroe and
Johnson.}
\end{figure}

\section{Future Probes at Auger}
\label{sec:Auger}

\begin{figure}[tbp]
\begin{minipage}[t]{0.49\textwidth}
\postscript{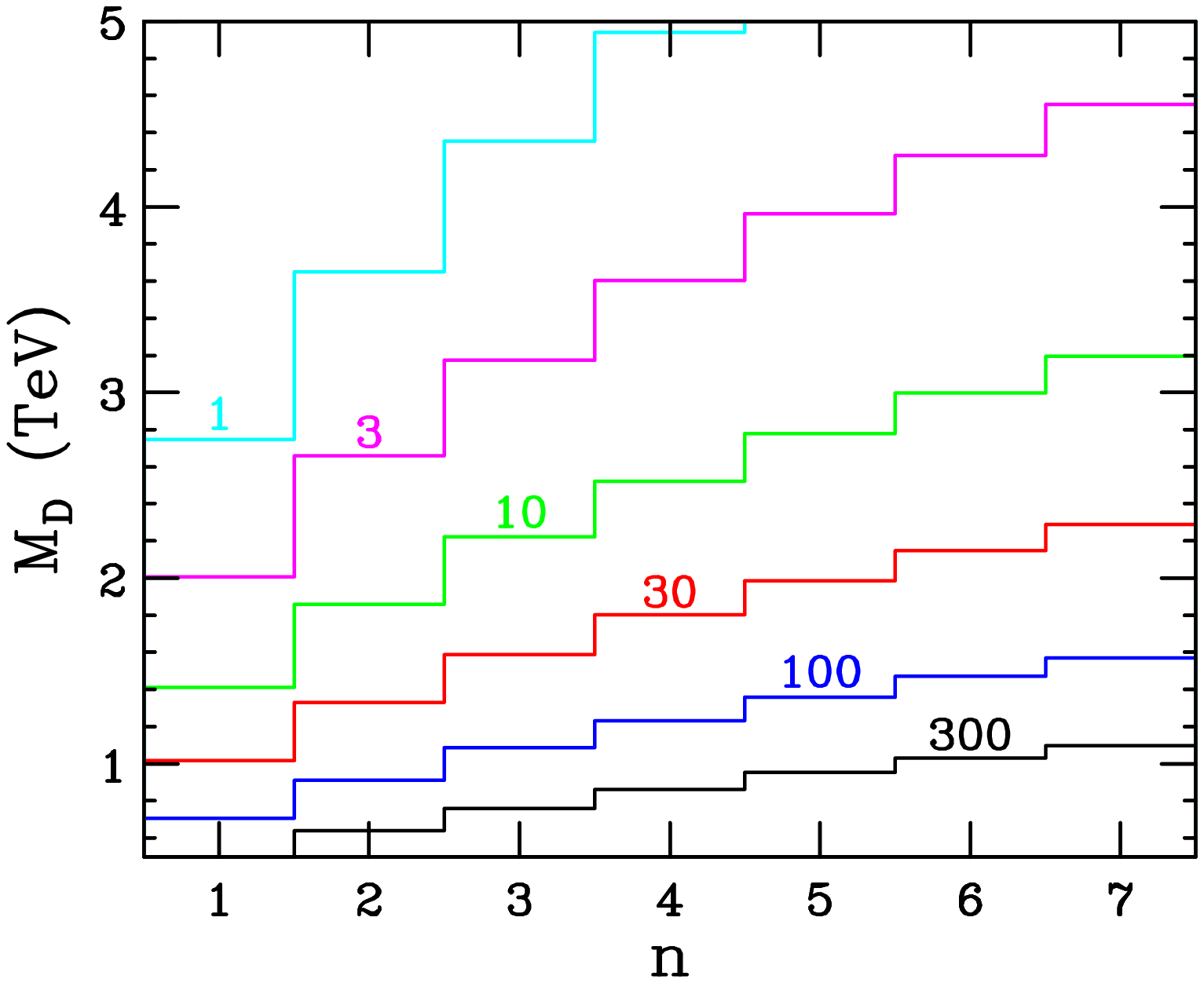}{0.99}
\end{minipage}
\hfill
\begin{minipage}[t]{0.49\textwidth}
\postscript{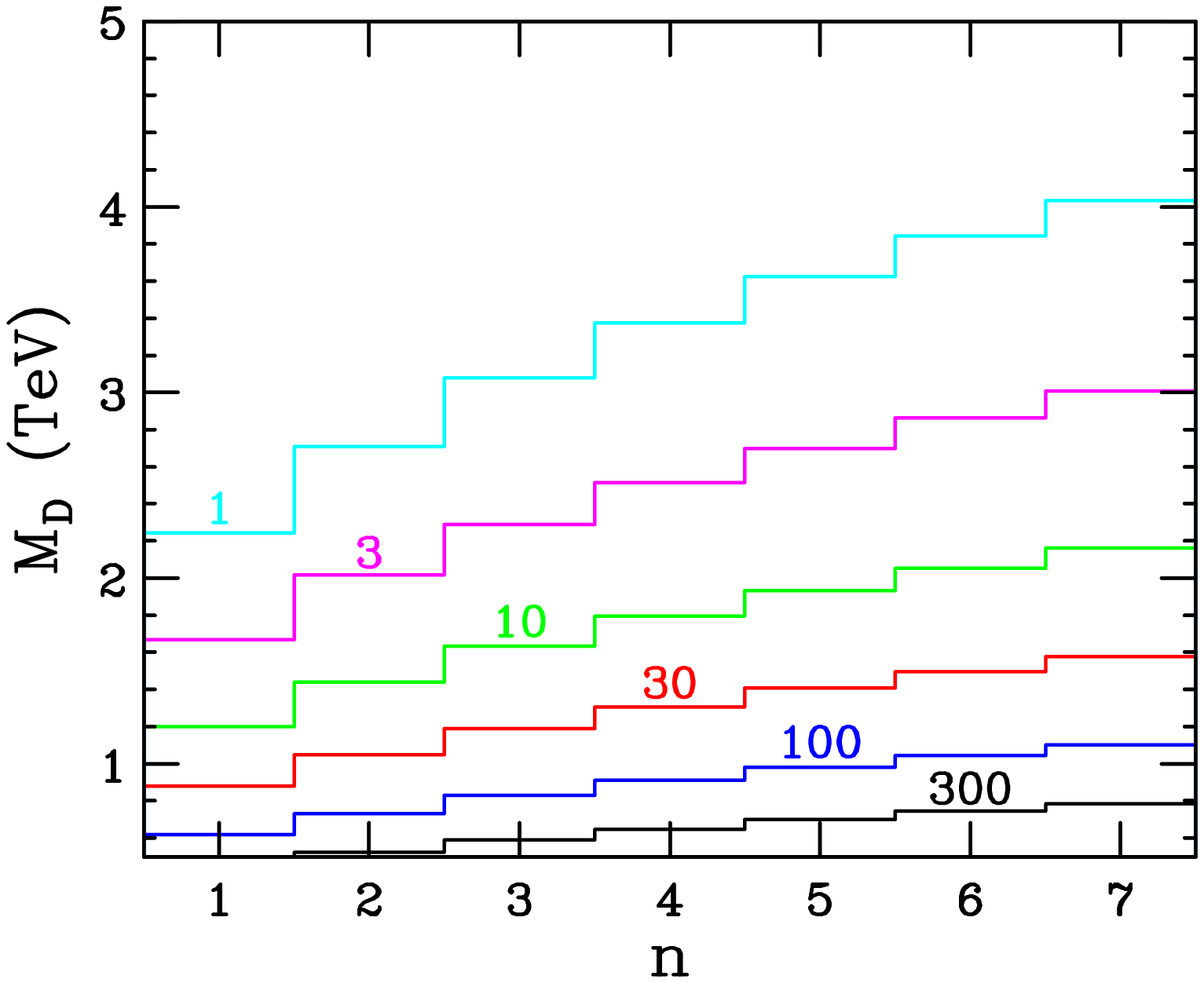}{0.99}
\end{minipage}
\caption{Event rates in 5 years for the Auger ground array for
$\xmin=1$ (left) and 3 (right).}
\label{fig:Auger_n_MD_1}
\end{figure}

Given the apertures discussed in Sec.~\ref{sec:acceptance}, it is a
simple matter to estimate the BH event rate for Auger.  The number of
detected BH events are given in Fig.~\ref{fig:Auger_n_MD_1} for
various $n$ as a function of $\md$.  The Auger ground array is
expected to become fully operational in 2003.  We assume a running
time of 5 years, roughly the data expected before the LHC begins.  For
$\xmin = 1$, Auger will probe fundamental Planck scales as large as
$\md = 4~\tev$.  For $\md \approx 1~\tev$ and $n\ge 4$, 100 BHs could
be detected.

\begin{figure}[tbp]
\postscript{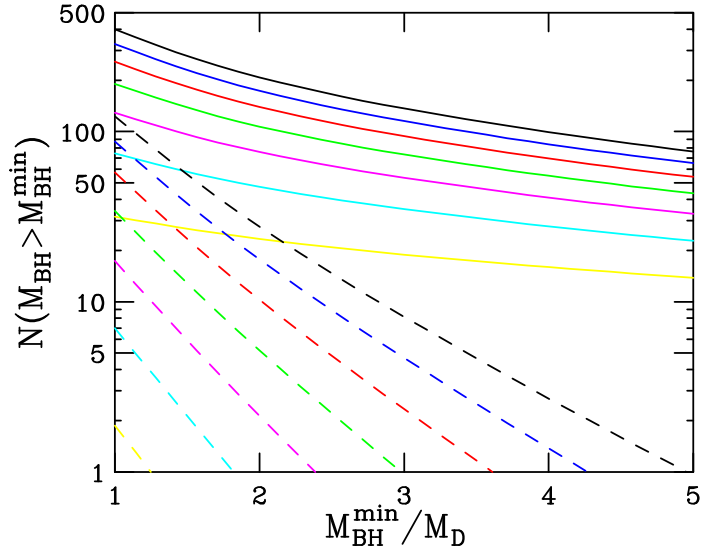}{0.56}
\caption{Event rates for BHs with mass above $\mbhmin$ at the
Auger ground array as a function of $\mbhmin/\md$ for $n=1,\ldots, 7$
from below, assuming $\md = 1~\tev$, 5 years running time, and parton
cross section $\pi r_s^2$ (solid) and $\pi r_s^2 e^{-I_E}$ (dashed).}
\label{fig:dndmbh}
\end{figure}

Given the prospects for fairly high statistics, detailed BH studies
are in principle possible.  While BHs with mass near $\md$ are in some
sense of the greatest interest, for detailed studies, one might first
restrict attention to more massive BHs (more energetic showers), where
the semi-classical description of BHs is expected to be justified.
The distribution of BH masses in cosmic ray collisions is given in
Fig.~\ref{fig:dndmbh}.  They are concentrated near $\md$, but the
event rate is reduced by only ${\cal O}(1)$ factors for $\xmin$ as
large as 5.  This contrasts strongly with the case at colliders, where
there is little energy to spare, cross sections are suppressed by two
parton distribution functions, and event rates are reduced by two
orders of magnitude for $\xmin = 5$ relative to $\xmin =
1$~\cite{Dimopoulos:2001hw}.  Total event rates for $\xmin = 3$ are
also given in Fig.~\ref{fig:Auger_n_MD_1}.  Even for $\xmin = 3$, we
find that $\sim 100$ BHs may be detected for $\md$ near $1~\tev$.

\begin{figure}[tbp]
\begin{minipage}[t]{0.49\textwidth}
\postscript{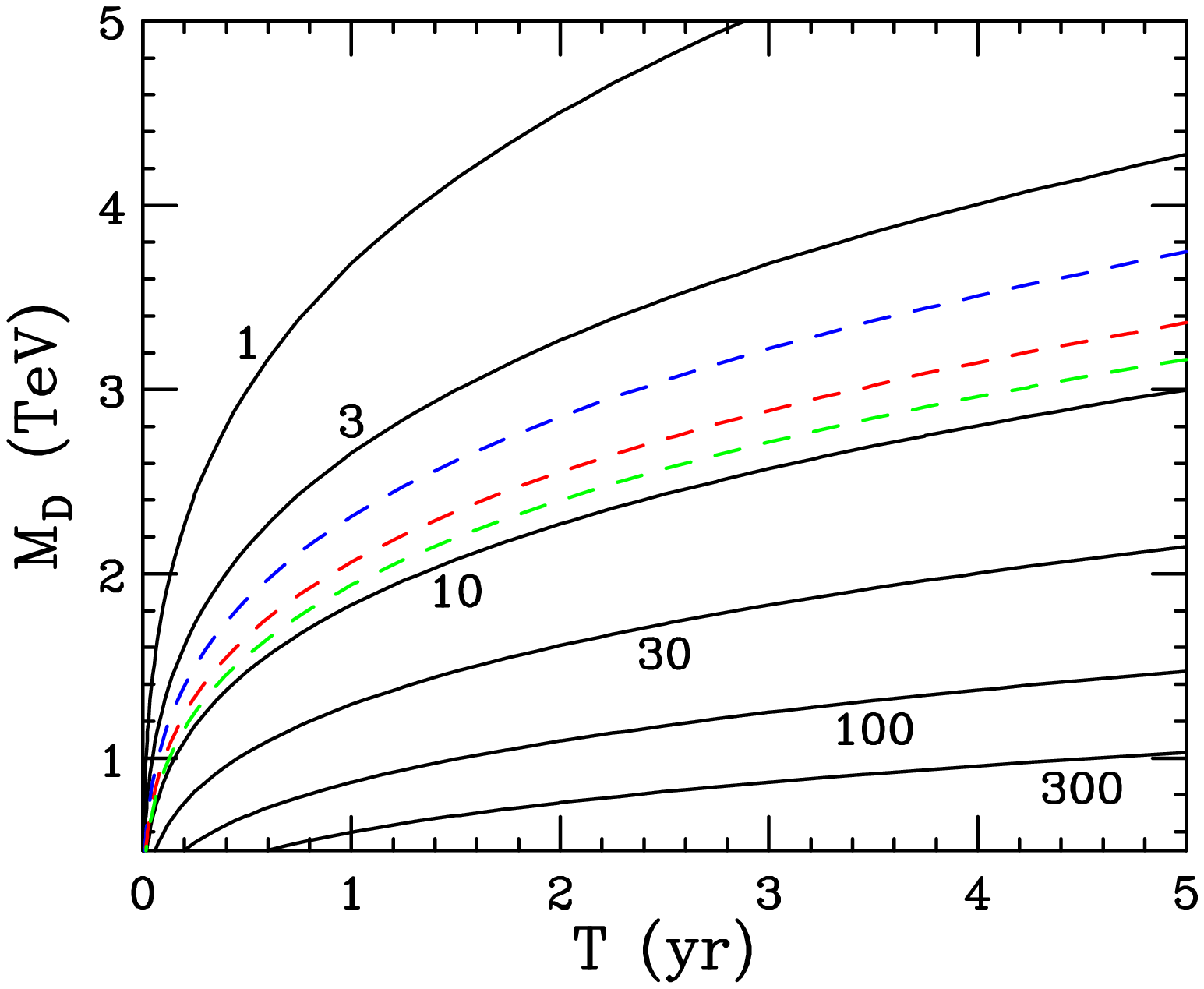}{0.99}
\end{minipage}
\hfill
\begin{minipage}[t]{0.49\textwidth}
\postscript{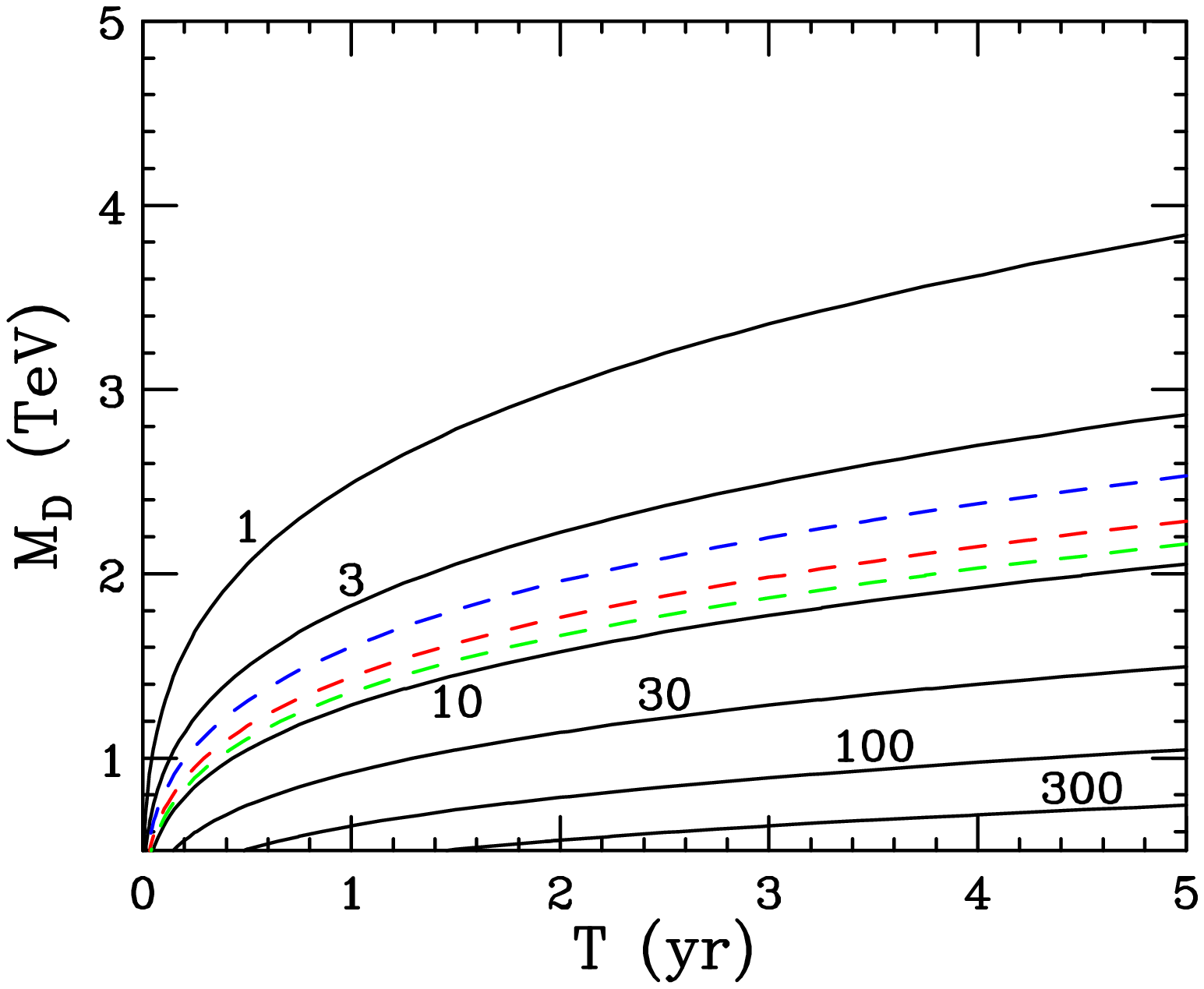}{0.99}
\end{minipage}
\caption{BH event rates at the Auger ground array for $n=6$ and
$\xmin=1$ (left) and 3 (right). The dashed contours indicate the
expected 95\% CL lower bound on $\md$ in the absence of physics beyond
the SM and assume $n_B = 0$, 5, and 10 background events from hadronic
showers (from above). The geometric cross section $\pi r_s^2$ is
assumed.}
\label{fig:Auger_T_MD_1}
\end{figure}

The dependence of BH event rates on running time $T$ is given in
Fig.~\ref{fig:Auger_T_MD_1}.  The event rate contours rise rapidly at
first --- in even the first few months, Auger will be sensitive to
values of $\md$ beyond present experiments.

If no enhancement of quasi-horizontal showers is seen, Auger will set
stringent limits on low-scale gravity and scenarios with extra
dimensions.  To determine these limits, we again assume the cosmogenic
fluxes of Sec.~\ref{sec:flux} and that only SM sources of deeply
penetrating showers are observed.  In contrast to AGASA, SM neutrino
interactions lead to observable rates -- given the cross section of
Fig.~\ref{fig:sigma}, 0.5 events per year are expected.  In addition,
as at AGASA, hadronic showers may fake deeply penetrating showers.  As
noted above, the Auger aperture of Sec.~\ref{sec:acceptance} assumes
zenith angles $\theta > 75^{\circ}$, corresponding to slant depths of
$\xmax \agt 4000~\g/\cm^2$, significantly more stringent than for the
AGASA study~\cite{agasa}.  Nevertheless, hadronic showers may be a
significant background.  We know of no detailed study, but consider
the possibility of $n_B$ background events from hadronic showers in 5
years below.

Given these assumptions, the expected background in 5 years is roughly
$2+n_B$ events.  To determine the expected limit on BH production we
assume that $2+n_B$ deeply penetrating events are in fact observed.
At 95\% CL, then, the upper bound on signal events for $n_B = 0$, 5,
and 10 is 4.7, 6.8, and 8.3 events, respectively.  in
Fig.~\ref{fig:Auger_T_MD_1}, contours for these event rates are also
given.  We find that, for $n_B \le 10$, $\xmin = 1$, and $n=6$, if no
events above background are observed, Auger will extend current bounds
on $\md$ to above 2 TeV after the first year of live time.  After 5
years, for $\xmin=1$, Auger will set a limit of $\md \agt 3~\tev$ for
$n \ge 4$.  In conjunction with astrophysical bounds, this will
require $\md \agt 3~\tev$ for all $n$, significantly straining
attempts to identify the Planck scale with the weak scale in scenarios
of large extra dimensions.  Note that we have neglected
model-dependent sub-Planckian effects that may increase the rates and
strengthen the bounds presented here.

\begin{figure}[tbp]
\begin{minipage}[t]{0.49\textwidth}
\postscript{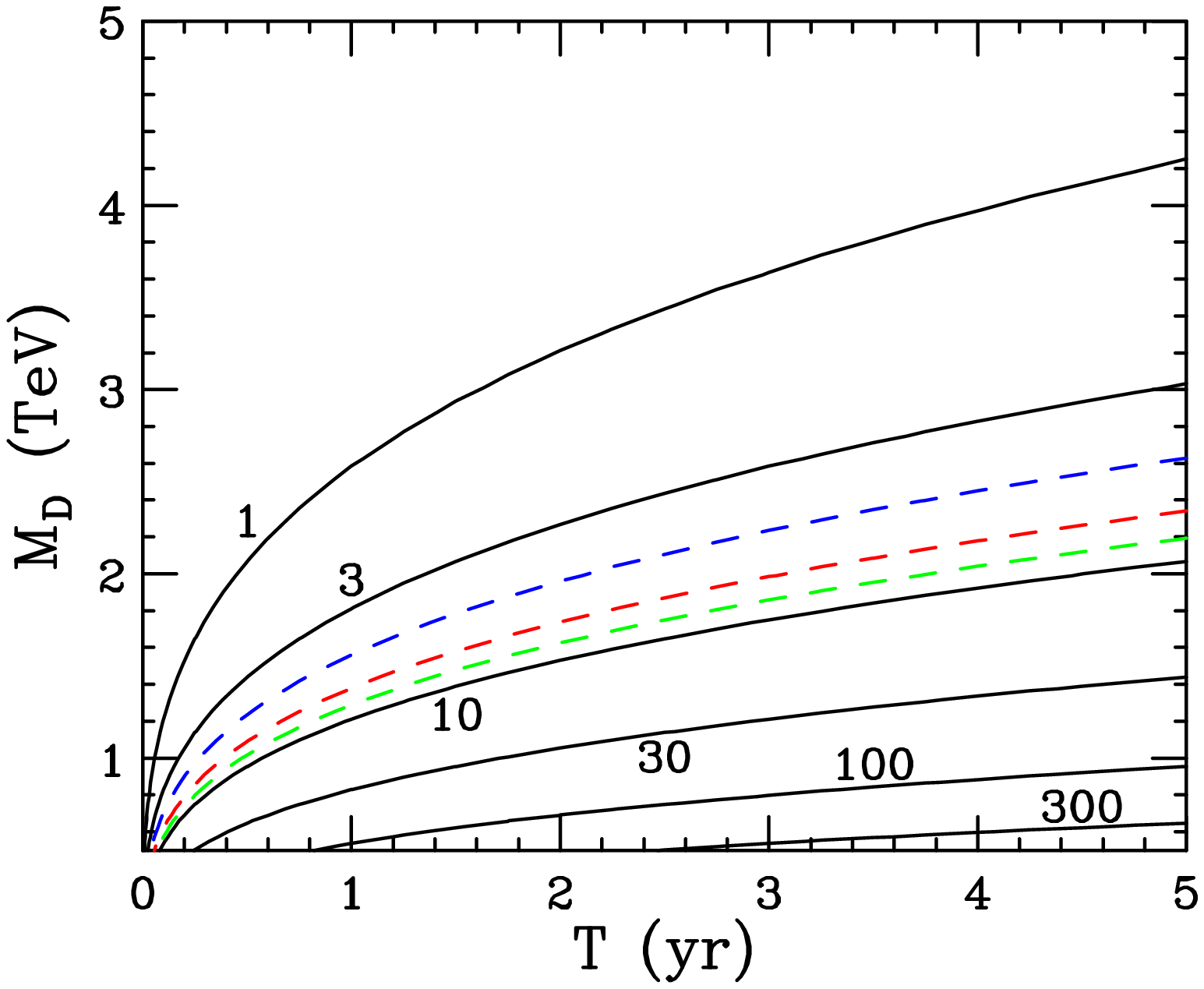}{0.99}
\end{minipage}
\hfill
\begin{minipage}[t]{0.49\textwidth}
\postscript{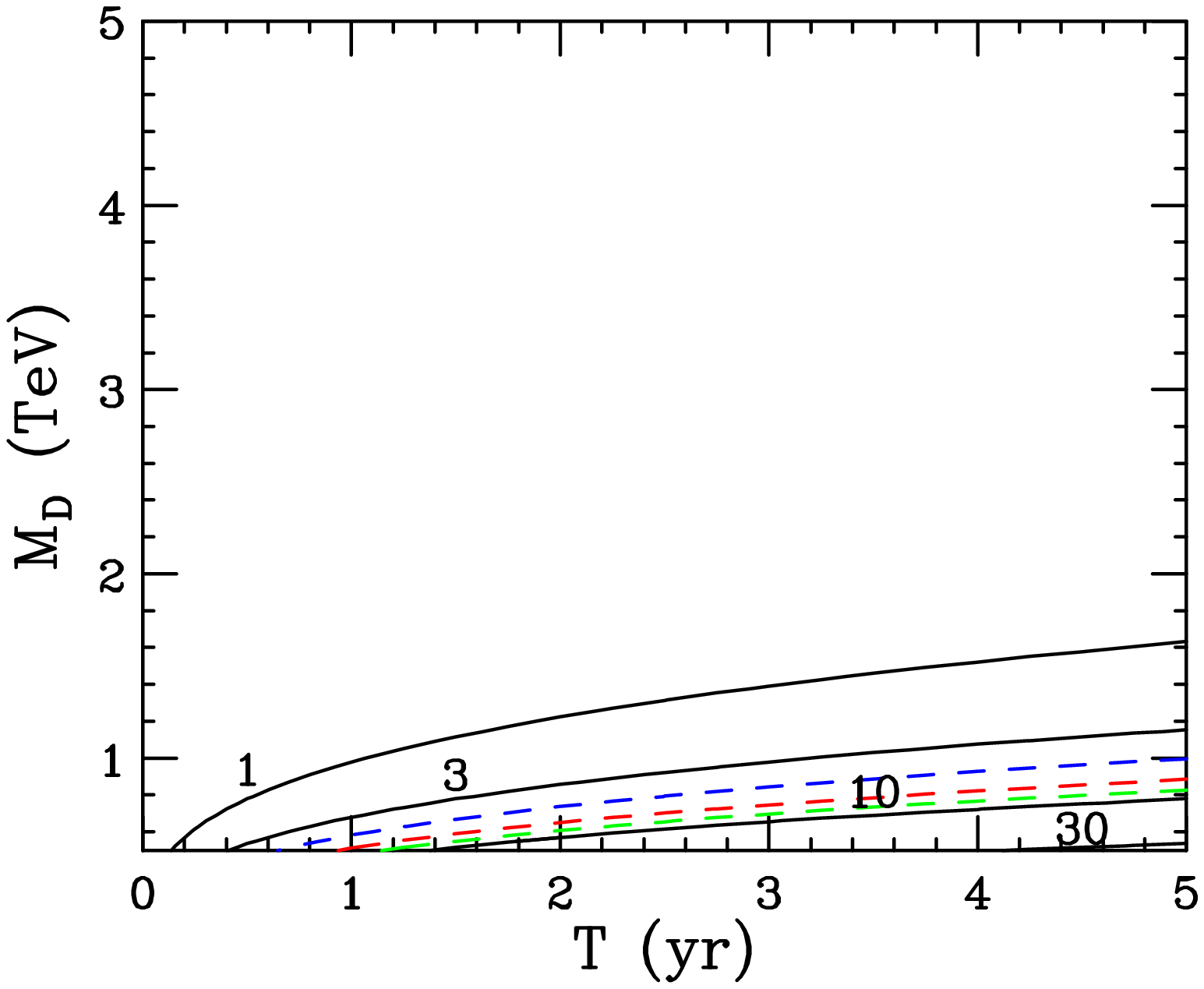}{0.99}
\end{minipage}
\caption{BH vent rates as in Fig.~\protect\ref{fig:Auger_T_MD_1}, but
for parton cross section $\pi r_s^2 e^{-I_E}$.}
\label{fig:Auger_T_MD_V1}
\end{figure}

Finally, we consider the impact of the proposed exponential
suppression of BH production cross sections.  In
Fig.~\ref{fig:dndmbh}, we show the dependence on $\xmin= \mbhmin/M_D$
for black hole event rates including this suppression.  For $\xmin =
1$, the exponential suppression is not particularly severe, reducing
event rates by factors of 3 for large $n$.  Of course, the impact is
much larger for larger $\xmin$.  In Fig.~\ref{fig:Auger_T_MD_V1}, we
show the number of BHs observed in time $T$ for parton cross section
$\pi r_s^2 e^{-I_E}$.  For $\xmin = 1$, Auger may still see tens of
BHs in 5 years, and will extend current bounds to $\md \approx
2.5~\tev$.  For $\xmin = 3$, the event rates are quite suppressed, but
a few BH events are still observable in 5 years.

\section{Distinguishing Black holes from SM events with Earth-skimming
neutrinos}
\label{sec:skimming}

If an excess of quasi-horizontal showers is observed, how can it be
identified as arising from BH events?  After all, at first sight, an
excess may arise simply from an enhanced flux.  With sufficient
statistics, a SM explanation may be excluded based on shower
properties, as black hole showers differ markedly from those produced
by SM charged and neutral current neutrino
interactions~\cite{Anchordoqui:2001ei}.  It may also be possible to
confirm specific predictions of BH production by verifying Hawking
radiation through correlations between $\xmax$ and shower
energy~\cite{Feng:2001ib}.

It is also possible, however, to differentiate BH from SM events by
considering additional constraints on ultra-high energy neutrino
properties.  In particular, comparison with Earth-skimming neutrino
rates may allow one to distinguish BH and SM
interpretations~\cite{Feng:2001ib}.  In this section, we develop this
possibility quantitatively, focusing on the question of excluding a SM
interpretation for BH events.  We also comment briefly on the task of
differentiating black hole events from other new physics possibilities
at the end of this Section.

\begin{figure}[tbp]
\postscript{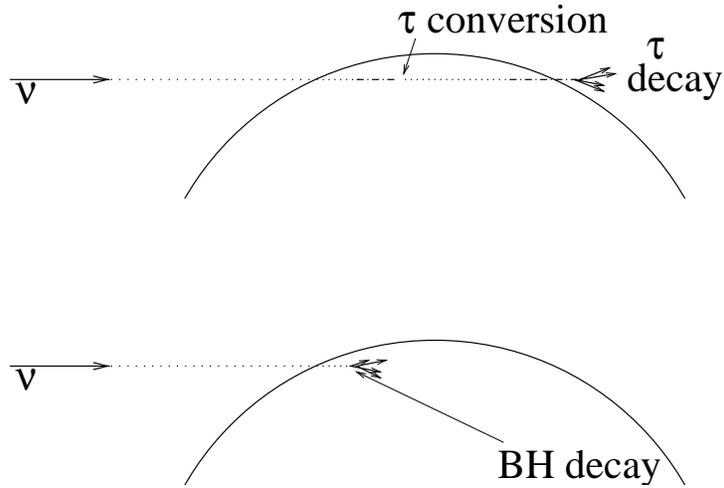}{0.66}
\caption{Top: A neutrino enters the Earth and converts into a charged
lepton, which exits the Earth and may be detected.  Bottom: A neutrino
enters the Earth and produces a BH, which is captured in the Earth.}
\label{fig:skimming}
\end{figure}

At ultra-high energies, even the SM neutrino cross section is large
enough that upward-going neutrinos are blocked by the Earth.  However,
neutrinos that skim the Earth, traveling at low angles along chords
with lengths of order their interaction length, are
not~\cite{Bertou:2001vm,Feng:2001ue,Kusenko:2001gj,Domokos:1997ve}. These
Earth-skimming neutrinos may then convert to charged leptons in the
Earth's crust, and the resulting charged leptons may emerge into the
atmosphere, producing a signal in cosmic ray detectors.  A schematic
picture of such an event is given in the top panel of
Fig.~\ref{fig:skimming}. The best signal is from $\tau$
leptons. Unlike electrons that do not escape the Earth's crust, or
muons that do not produce any visible signal in the atmosphere, taus
can travel for tens of km in rock, escape, and then decay in the
atmosphere, leading to spectacular showers and observable rates of
order 1 per year in both ground arrays~\cite{Bertou:2001vm} and
fluorescence detectors~\cite{Feng:2001ue}.  The optimal angle for
Earth-skimming neutrinos is energy-dependent.  For $E_{\nu} \sim
10^{8}\ (10^{10})~\gev$, the optimal angle relative to the horizon is
$\sim 3^{\circ} \ (1^{\circ})$.  Given the angular resolution of
cosmic ray detectors, these Earth-skimming events are easily
differentiated from standard horizontal neutrino showers.

The scenario changes radically in the presence of a significant cross
section for BH production.  First, BHs decay largely to hadrons, which
do not escape the Earth.  Such an event is pictured in the bottom
panel of Fig.~\ref{fig:skimming}. Of course, BHs also have a
significant leptonic branching fraction, but leptons from BH decay
carry only a fraction of the initial neutrino energy, and their
detection rate is therefore highly suppressed.  The probability of
detecting BHs produced in the Earth by ground arrays and surface
fluorescence detectors is therefore insignificant.  Second, a
sufficiently large BH cross section also depletes the original
neutrino beam through absorption, leading to a substantial suppression
of all Earth-skimming events, including those in the top panel of
Fig.~\ref{fig:skimming}.

To determine the effects of BH production on Earth-skimming rates, we
consider here a simple analysis that is nevertheless sufficient to
isolate the functional dependence of Earth-skimming rates on cross
section parameters.  The analysis extends the discussion of
Ref.~\cite{Feng:2001ue}, where additional details and discussion may
be found.

Consider a flux of neutrinos with energy $E_0$.  Given the high
energies required for detection, the most relevant energies are $E_0
\sim 10^9 - 10^{10}~\gev$, even for cosmogenic flux evaluations peaked
at somewhat lower energies, and we may therefore limit the discussion
to this rather narrow band of energy.  Earth-skimming events occur in
the Earth's crust, and so the relevant neutrinos and taus sample only
the Earth's surface density, $\rho_s \approx 2.65~\g/\cm^3$.  In the
SM, the neutrino's path length is
\begin{equation}
L_{\rm CC}^{\nu} = \left[ N_A \rho_s \sigma^{\nu}_{\rm CC}
\right] ^{-1} \ ,
\end{equation}
where $N_A \simeq 6.022 \times 10^{23}~\g^{-1}$ and
$\sigma^{\nu}_{\rm CC}$ is the charged current cross sections for
$E_{\nu} = E_0$.  (We neglect neutral current interactions, which at
these energies serve only to reduce the neutrino energy by
approximately 20\%.)  For $E_0 \sim 10^{10}~\gev$, $L_{\rm CC}^{\nu}
\sim {\cal O}(100)~\km$.  Supplemented by the possibility of BH
production, the neutrino's path length is
\begin{equation}
L_{\rm tot}^{\nu} = \left[ N_A \rho_s (\sigma^{\nu}_{\rm CC} +
\sigma^{\nu}_{\rm BH}) \right] ^{-1} \ ,
\end{equation}
where $\sigma^{\nu}_{\rm BH}$ is the BH production cross section
for $E_{\nu} = E_0$.

At these energies, the tau's propagation length is determined not by
its decay length but by its energy loss.  The $\tau$ lepton loses
energy in the Earth according to
\begin{equation}
\frac{dE_{\tau}}{dz} = -(\alpha_{\tau} + \beta_{\tau} E_{\tau})
\rho_s \ ,
\end{equation}
where, for these energies, $\alpha_{\tau}$ is negligible, and we take
$\beta_{\tau} \approx 0.8\times 10^{-6}~\cm^2/\g$~\cite{Dutta:2000hh}.
The maximal path length for a detectable $\tau$ is, then,
\begin{equation}
L^{\tau} = \frac{1}{\beta_{\tau} \rho_s} \ln \left( E_{\rm max} /
E_{\rm min} \right) \ ,
\label{ltau}
\end{equation}
where $E_{\rm max}\approx E_0$ is the energy at which the tau is
created, and $E_{\rm min}$ is the minimal energy at which a $\tau$ can
be detected.  For cosmogenic neutrino fluxes and other reasonable
sources, and the acceptances of typical cosmic ray detectors, taus
cannot lose much energy and be detected.  For $E_{\rm max} / E_{\rm
min} = 10$, $L^{\tau} = 11~\km$.

Given an isotropic $\nu_{\tau} + \bar{\nu}_{\tau}$ flux, the number of
taus that emerge from the Earth with sufficient energy to be detected
is proportional to an ``effective solid angle''
\begin{equation}
\Omega_{\rm eff} \equiv \int d\cos\theta\, d\phi\, \cos\theta\,
P(\theta,\phi) \ ,
\end{equation}
where
\begin{equation}
P(\theta,\phi) = \int_0^{\ell} \frac{dz}{L_{\rm CC}^{\nu}}
e^{-z/L_{\rm tot}^{\nu}} \
\Theta \left[ z - (\ell - L^{\tau} ) \right]
\label{P}
\end{equation}
is the probability for a neutrino with incident nadir angle $\theta$
and azimuthal angle $\phi$ to emerge as a detectable $\tau$. (In
\eqref{P}, for the reasons noted above, we have neglected the
possibility of detectable signals from BH production by Earth-skimming
neutrinos.)  Here $\ell = 2 R_{\oplus} \cos\theta$ is the chord length
of the intersection of the neutrino's trajectory with the Earth, with
$R_{\oplus} \approx 6371~\km$ the Earth's radius.  Evaluating the
integrals, we find~\cite{Kusenko:2001gj}
\begin{equation}
\Omega_{\rm eff} = 2 \pi
\frac{L_{\rm tot}^{\nu}}{L_{\rm CC}^{\nu}}
\left[ e^{L^{\tau} / L_{\rm tot}^{\nu}} - 1 \right]
\left[ \left( \frac{L_{\rm tot}^{\nu}}{2 R_{\oplus}} \right)^2
- \left( \frac{L_{\rm tot}^{\nu}}{2 R_{\oplus}} +
\left( \frac{L_{\rm tot}^{\nu}}{2 R_{\oplus}} \right)^2 \right)
e^{-2R_{\oplus} / L_{\rm tot}^{\nu}} \right] \ .
\label{Omegaeff}
\end{equation}
At the relevant energies, the neutrino interaction length satisfies
$L_{\rm tot}^{\nu} \ll R_{\oplus}$.  In addition, for $L_{\rm
tot}^{\nu} \gg L^{\tau}$, valid when the BH cross section is not very
large, \eqref{Omegaeff} simplifies to
\begin{equation}
\Omega_{\rm eff} \approx
2\pi \frac{L_{\rm tot}^{\nu\, 2} L^{\tau}}{4 R_{\oplus}^2
L_{\rm CC}^{\nu}} \ .
\label{omegaeffapprox}
\end{equation}

Equation~(\ref{omegaeffapprox}) gives the functional dependence of the
Earth-skimming event rate on the BH cross section.  This rate is, of
course, also proportional to the source neutrino flux $\Phi^{\nu}$ at
$E_0$.  Finally, the constant of proportionality is determined by
previous studies~\cite{Bertou:2001vm,Feng:2001ue}, where all the
experimental issues entering tau detection have been included.  Given
these inputs, the number of Earth-skimming neutrino events detected in
5 years is
\begin{equation}
N_{\rm ES} \approx
C_{\rm ES}\, \frac{\Phi^{\nu}}{\Phi^{\nu}_0}
\frac{\sigma^{\nu\, 2}_{\rm CC}}
{\left( \sigma^{\nu}_{\rm CC} + \sigma^{\nu}_{\rm BH} \right)^2} \ ,
\label{ES}
\end{equation}
where $C_{\rm ES}$ is the number of Earth-skimming events expected for
the standard cosmogenic flux $\Phi^{\nu}_0$ in the absence of BH
production.  For detection by the Auger ground array, $C_{\rm ES}
\approx 3.0$, assuming maximal neutrino mixing and the $\beta_{\tau}$
value given above~\cite{Bertou:2001vm}.  The fluorescence detectors of
HiRes provide additional sensitivity~\cite{Feng:2001ue}, as do those
of Auger~\cite{Cester}.  We conservatively take $C_{\rm ES} = 3$ for
the combined rate in 5 years expected in the SM.  Note, however, that
the rate may be greatly suppressed for large BH cross sections, as
anticipated.

In contrast to \eqref{ES}, the rate for quasi-horizontal showers
follows simply from \eqref{numevents}, and has the form
\begin{equation}
N_{\rm QH} = C_{\rm QH} \frac{\Phi^{\nu}}{\Phi^{\nu}_0}
\frac{\sigma^{\nu}_{\rm CC} + \sigma^{\nu}_{\rm BH}}
{\sigma^{\nu}_{\rm CC}} \ ,
\end{equation}
where $C_{\rm QH} = 2.5$ for the Auger ground array, as noted
previously.

\begin{figure}[tbp]
\begin{minipage}[t]{0.49\textwidth}
\postscript{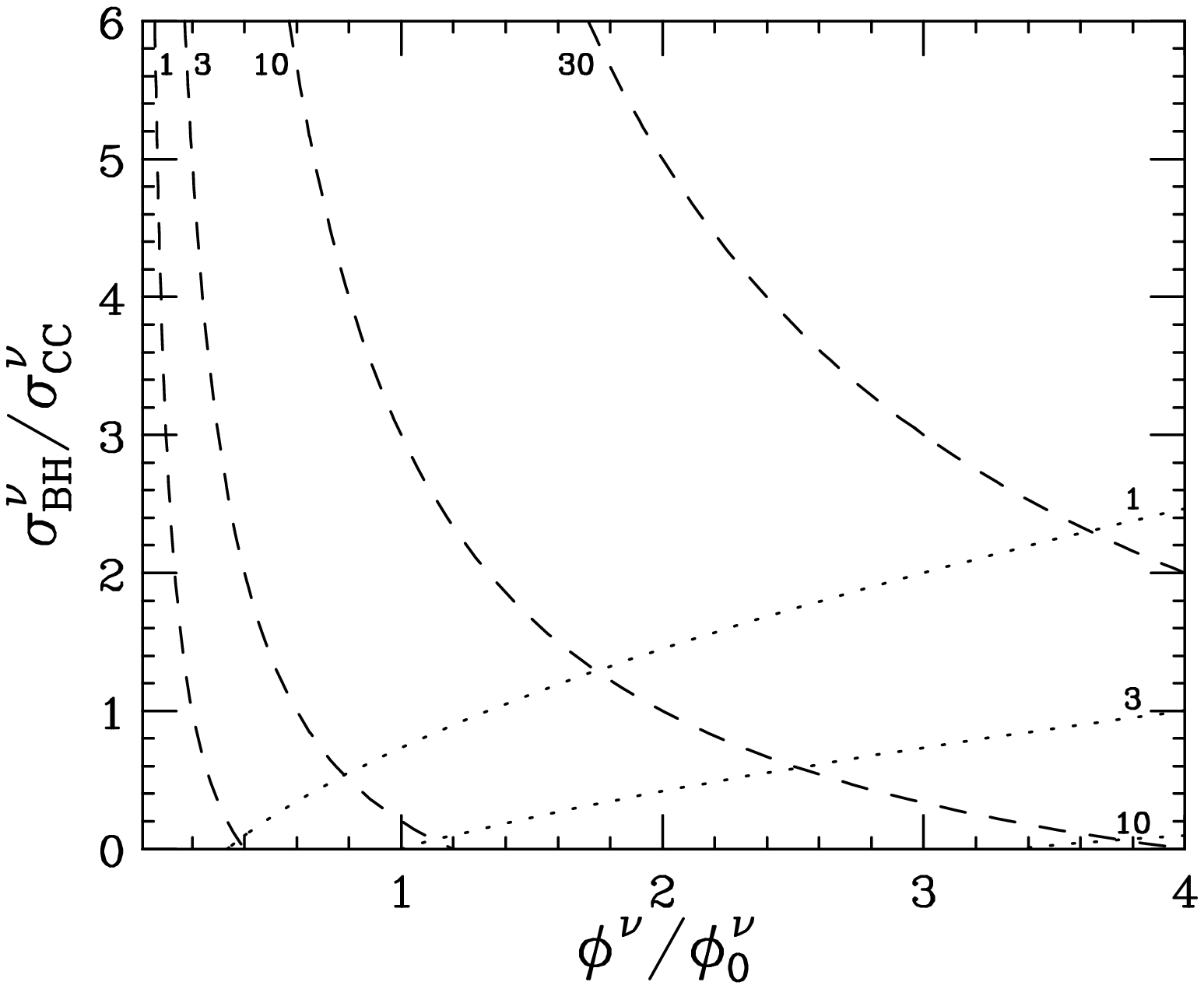}{0.99}
\end{minipage}
\hfill
\begin{minipage}[t]{0.49\textwidth}
\postscript{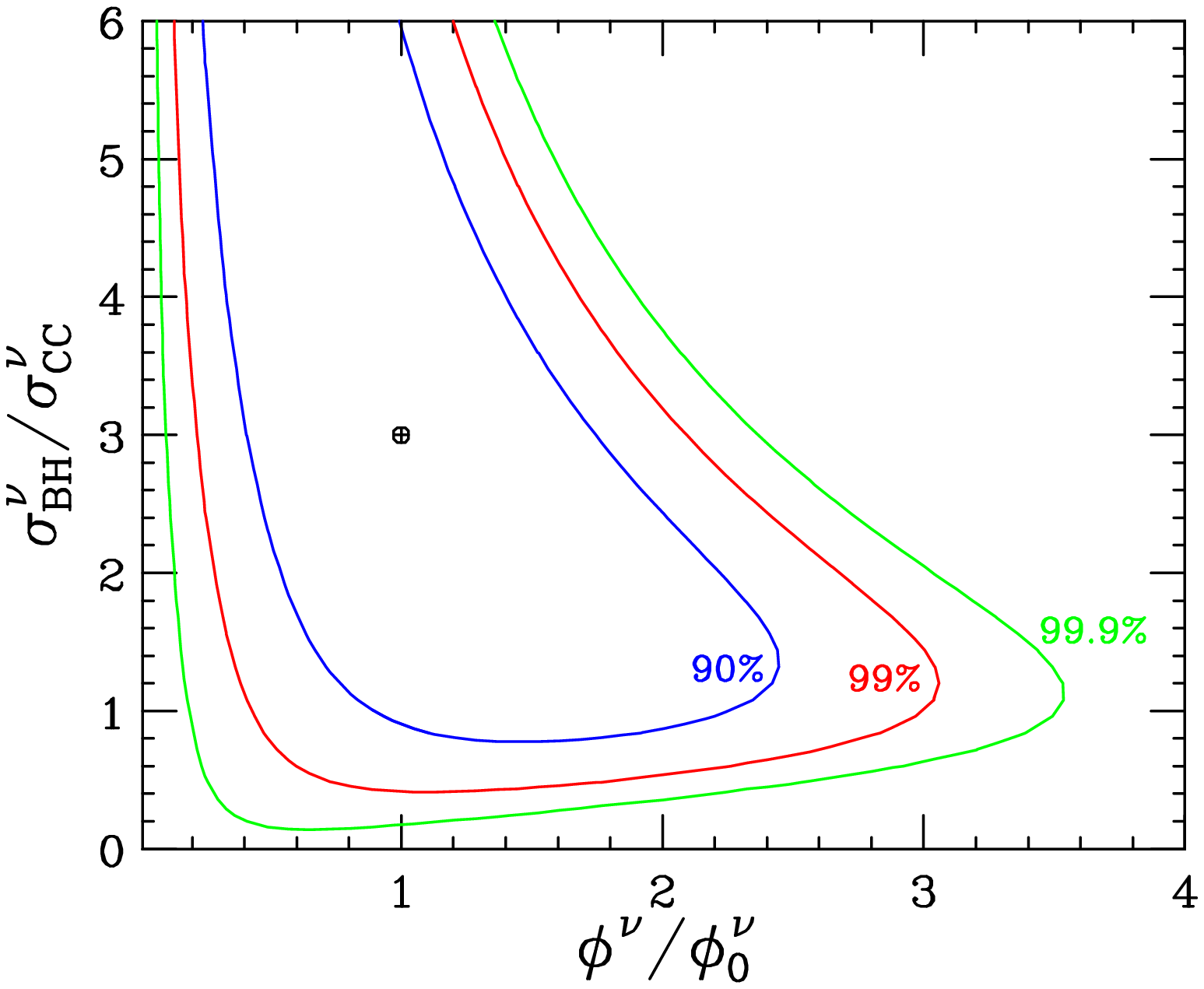}{0.99}
\end{minipage}
\caption{Left: Contours of constant number of quasi-horizontal showers
$N_{\rm QH}$ (dashed) and Earth-skimming neutrino events $N_{\rm ES}$
(dotted) as functions of source flux $\Phi^{\nu}$ and BH production
cross section $\sigma^{\nu}_{\rm BH}$.  5 year running times for Auger
and HiRes are assumed. Right: Confidence level contours, assuming
$\Phi^{\nu} = \Phi^{\nu}_0$ and $\sigma^{\nu}_{\rm BH} = 3
\sigma^{\nu}_{\rm CC}$, corresponding to $(N_{\rm QH}, N_{\rm ES})
\approx (10, 0.2)$.  }
\label{fig:skimbh}
\end{figure}

Given a flux $\Phi^{\nu}$ and BH cross section $\sigma^{\nu}_{\rm
BH}$, both $N_{\rm ES}$ and $N_{\rm QH}$ are determined.  Event
contours are given in the left panel of Fig.~\ref{fig:skimbh}. As can
be seen, given a quasi-horizontal event rate $N_{\rm QH}$, it is
impossible to differentiate between an enhancement from large BH cross
section and large flux.  However, in the region where significant
event rates are expected, the $N_{\rm QH}$ and $N_{\rm ES}$ contours
are more or less orthogonal, and provide complementary information.
With measurements of $N_{\rm QH}$ and $N_{\rm ES}$, both
$\sigma^{\nu}_{\rm BH}$ and $\Phi^{\nu}$ may be determined
independently, and neutrino interactions beyond the SM may be
unambiguously identified. (See also Ref.~\cite{Kusenko:2001gj}.)

As an example, consider the case in which $\sigma^{\nu}_{\rm
BH}/\sigma^{\nu}_{\rm CC} = 3$, and $\Phi^{\nu}/\Phi^{\nu}_0 = 1$. On
average, one would then observe a total of $N_{\rm QH} = 10$ deep
quasi-horizontal showers, an excess of 8 above SM expectations.  On
average, one also expects $N_{\rm ES} \approx 0.2$ Earth-skimming
events.  A SM explanation (with $\sigma^{\nu}_{\rm BH} = 0$) of the
deeply penetrating event rate would require $\Phi^{\nu}/\Phi^{\nu}_0 =
4$ and predict 12 Earth-skimming events, a possibility that would be
clearly excluded at high confidence level.

More generally, one might try to salvage a SM explanation by
attributing the observed rates to statistical fluctuations in both
$N_{\rm QH}$ and $N_{\rm ES}$.  Using a maximum likelihood method for
Poisson-distributed data~\cite{James:sr}, we give contours of constant
$\chi^2$ in the right panel of Fig.~\ref{fig:skimbh}.  The possibility
of a SM interpretation along the $\sigma^{\nu}_{\rm BH} = 0$ axis
would be excluded at greater than 99.9\% CL for any assumed flux.  The
power of the Earth-skimming information is such that the best fit is
in fact found for $\Phi^{\nu} < \Phi^{\nu}_0$!  We find, then, that if
even an excess of a handful of quasi-horizontal events is observed, by
comparing to the Earth-skimming neutrino rate, attempts to explain the
excess by SM interactions alone may be excluded. These arguments
require only counting experiments, and do not rely on measurements of
shower properties.

BH production will most likely be accompanied by more
model-independent sub-Planckian effects.  In particular, neutral
current neutrino cross sections may be enhanced in extra-dimensional
scenarios through the exchange of KK gravitons.  This will raise the
quasi-horizontal rate, but will have very little effect on the
Earth-skimming event rate, since neutrinos suffer very little energy
loss during this process~\cite{Emparan:2001kf}.  We expect such
effects, then, to further enhance the ratio $N_{\rm QH}/N_{\rm ES}$,
making a SM explanation even more untenable.

So far, we have not explicitly considered the question of
distinguishing BH events from other types of new physics.  However,
the prediction of enhanced quasi-horizontal event rates {\em and}
diminished Earth-skimming rates is incisive.  For example, new physics
that increases quasi-horizontal rates by enhancing cross sections for
$\nu N \to \ell X$ will also increase Earth-skimming rates.  The
prediction of suppressed Earth-skimming rates relies on the efficient
conversion of neutrino energy directly to hadronic energy, that is, a
process with large cross section and large inelasticity. This is a
peculiar property of BHs that separates BH production from other
possible forms of new physics.  The comparison between deep
quasi-horizontal shower and Earth-skimming neutrino rates therefore
not only effectively excludes a SM interpretation of BH events, but
goes a long way toward excluding other new physics explanations.

\section{Summary of Results and Conclusions}
\label{sec:conclusions}

In this work we have shown that cosmic ray observations in the recent
past (AGASA) and in the near future (Auger) provide extremely
sensitive probes of low-scale gravity and extra dimensions.  We have
focused on the production of TeV-scale BHs resulting from collisions
of ultra-high energy cosmic neutrinos in the Earth's atmosphere, and
have considered the impact of various theoretical issues in the
determination of the BH production cross section.  In particular, mass
shedding, the production of BHs with non-zero angular momentum, and a
possible enhancement of the BH cross section can be expected to give
minor perturbations.  The exponential suppression proposed by Voloshin
is more significant, but large and observable BH event rates are still
possible.

More specifically, in the case of $n$ extra spatial dimensions
compactified on an $n$-torus with a common radius, we have found the
following:

\begin{itemize}

\item{Present bounds on atmospheric BH production imply 95\% CL lower
limits on the fundamental Planck mass of $\md\ge 1.3-1.5~\tev$ for
$n=4$, rising to $\md\ge 1.6-1.8~\tev$ for $n=7$.  These bounds follow
from the non-observation of a significant excess of deep,
quasi-horizontal showers in 1710.5 days of running recently reported
by the AGASA Collaboration~\cite{agasa}.

The absence of a deeply-penetrating signal in the Fly's Eye
data~\cite{Baltrusaitis:mt} also implies lower bounds on $M_D$.  These
are consistently weaker, however.  For example, for $n=6$, $\xmin=1$,
and the same (PJ) flux we have used, Ringwald and Tu find $M_D >
900~\gev$~\cite{Ringwald:2001vk}. We find this difference to be
significant: the AGASA and Fly's Eye constraints rely on identical
theoretical assumptions, and given the scaling in \eqref{scaling}, a
factor of 2 difference in $\md$ bounds corresponds to a factor of more
than 4 in acceptance or, equivalently, running time.

The AGASA limits derived here exceed the D\O\ bound $\md\agt 0.6-1.2$
TeV, where the variation reflects uncertainty from the choice of
ultraviolet cutoff for graviton momenta transverse to the brane.  The
cosmic ray limits are subject to a separate set of uncertainties,
discussed at length above, but follow from conservative evaluations of
the neutrino flux and experimental aperture, and $\xmin=1$.  For
$\xmin=3$, these limits are somewhat reduced, but still generally
exceed the Tevatron bounds.

The cosmic ray bounds from AGASA therefore represent the best existing
limits on the scale of TeV-gravity for $n\ge 4$ extra spatial
dimensions.  A summary of the most stringent present bounds on $\md$
for $n \ge 2$ extra dimensions is given in Fig.~\ref{fig:summary}. }

\begin{figure}[tbp]
\postscript{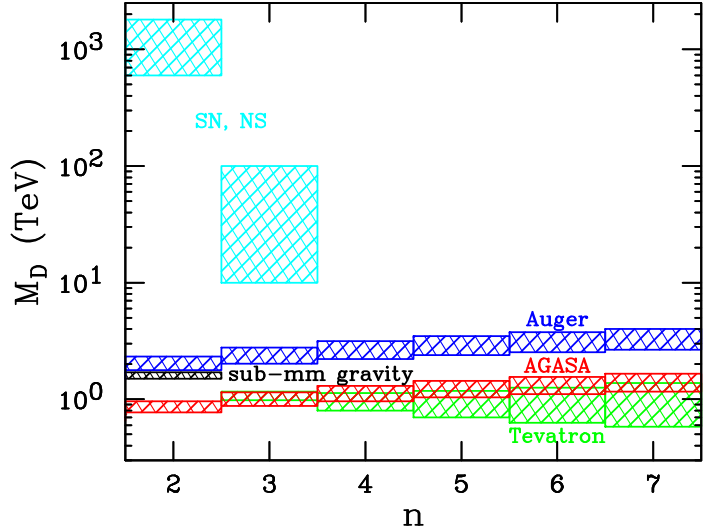}{0.56}
\caption{Bounds on the fundamental Planck scale $\md$ from tests of
Newton's law on sub-millimeter scales, bounds on supernova cooling and
neutron star heating, dielectron and diphoton production at the
Tevatron, and non-observation of BH production at AGASA.  Future
limits from the Auger ground array, assuming 5 years of data and no
excess above the SM neutrino background, are also shown.  The range in
Tevatron bounds corresponds to the range of brane softening parameter
$\Lambda/\md =0.5-1$.  The range in cosmic ray bounds is for
$\xmin=1-3$.  See text for discussion. }
\label{fig:summary}
\end{figure}

\item{The reach of AGASA will be extended significantly by the Auger
Observatory. If no quasi-horizontal extended air shower events are
observed in 5 years (beyond the expected two SM neutrino events
supplemented by as many as 10 hadronic background events), Auger will
set a limit of $\md\agt 3~\tev$, at 95\% CL, for $n \ge 4$.  Even in
the case where the cross section is decreased by the exponential
suppression factor in \eqref{sigmasupp}, a bound $\md\agt$ 2 TeV may
be found under the same background assumptions. }

\item{Conversely, given the large reach of Auger, tens of BH events
may be observed per year.  We have discussed in some detail how
combined measurements of quasi-horizontal air showers and
Earth-skimming $\nu_{\tau}\rightarrow \tau$ events may be used to
identify new neutrino interactions beyond the SM, even with complete
uncertainty about the incident neutrino flux.  In the case of BH
production, the quasi-horizontal event rate is enhanced, while the
Earth-skimming rate is suppressed, since BH production in the Earth
acts as an absorptive channel, depleting the SM rate. With counting
experiments alone, one can therefore exclude a SM interpretation of BH
events, and may distinguish BH events from almost all other possible
forms of new physics.}

\end{itemize}

In conclusion, in the next several years prior to the analysis of data
from the LHC, super-Planckian BH production from cosmic rays provides
a promising probe of extra dimensions.  Searches for BH-initiated
quasi-horizontal showers in the Earth's atmosphere at AGASA provide
the most stringent bounds on low-scale gravity at present, and the
Auger Observatory will extend this sensitivity to fundamental Planck
scales well above the TeV scale.

\begin{acknowledgments}
We thank R.~Emparan, S.~Giddings, A.~Ringwald, and G.~Sigl for useful
communications. JLF thanks A.~Guth and A.~Vilenkin for discussions
regarding Kerr BHs. ADS thanks D.~Eardley for a discussion about BH
collisions.  The work of LAA and HG has been partially supported by
the US National Science Foundation (NSF), under grants No.\
PHY--9972170 and No.\ PHY--0073034, respectively. The work of JLF was
supported in part by the Department of Energy (DOE) under cooperative
research agreement DF--FC02--94ER40818. The work of ADS is supported
in part by DOE Grant No.\ DE--FG01--00ER45832 and NSF Grant No.\
PHY--0071312.
\end{acknowledgments}


\end{document}